\def\kvec{ {\bf k}}
\def\Qvec{ {\bf Q}}

\def\kktil{\widetilde {k}}

\def\QQtil{\widetilde  { Q}}
\def\qtil{\widetilde {q}}

\def\GGtil{\widetilde {G}}

\def\0til{\bar {0}}

\def\dm{\frac{1}{2}}
\def\oq{\frac{1}{4}}
\def\sumslashD{\mathop{\sum \kern-1.4em -\kern 0.5em}}
\def\sumslash{\mathop{\sum \kern-1.2em -\kern 0.5em}}
\def\intslash{\mathop{\int \kern-0.9em -\kern 0.5em}}
\def\intslashD{\mathop{\int \kern-1.1em -\kern 0.5em}}

\documentclass{report}
\usepackage{svcon2e}
\usepackage{graphicx}
\begin{document}
\pagenumbering{arabic}

\bibliographystyle{prsty}

\chapter{Renormalization Group Technique for Quasi-one-dimensional Interacting Fermion Systems at Finite Temperature}
\chapterauthors{C. Bourbonnais, B. Guay and R. Wortis}

\begin{abstract}
   We review some aspects of the renormalization group method for interacting
fermions.   Special emphasis is
placed on the application of scaling theory to quasi-one-dimensional
systems at non zero temperature. We begin by
introducing the scaling ansatz for purely one-dimensional fermion systems  and
its extension when
interchain coupling and dimensionality
crossovers are present at  finite temperature. Next, we review
the application of the
renormalization  group  technique to the one-dimensional electron gas
model and clarify some  peculiarities  of
the method at the two-loop level. The influence of interchain coupling is
then included and results for   the
crossover phenomenology and the multiplicity of characteristic energy scales   are
summarized. The emergence of the   Kohn-Luttinger
mechanism in quasi-one-dimensional electronic structures is discussed for
both superconducting and density-wave
channels.
\end{abstract}

\tableofcontents

\section{Introduction}
Scaling ideas have  exerted a far-reaching influence on our understanding of complex many-body systems.
Their use in the study of critical phenomena has offered a broad and fertile field of work at the heart of
which is the Wilson formulation of the renormalization group method.
\cite{Wilson75,Wilson83} In the Wilson view, fluctuations of the order parameter  at all  length scales up to
the correlation length are the key ingredient in  explaining the  existence of  power law singularities that
govern critical properties  in accordance with    scaling laws. Equally powerful is the extension
of scaling concepts to anisotropic systems, namely when small anisotropic parameters are magnified as a
result of their coupling to singular fluctuations. Intermediate
 length scales thus emerge  giving  rise    to 
changes or {\it crossovers} in the critical behavior.\cite{Riedel69,Pfeuty74,Grover73} The horizon of
applications of this  methodology  was further widened  when scaling was applied to quantum critical
systems  for which quantum mechanics governs fluctuations of the order parameter at the critical
point.\cite{Hertz76,Sachdev99} 

There has been a parallel expansion of the use of    scaling tools in the description of  {\it
many-fermion} systems. This was especially true for the Kondo
impurity  and the one-dimensional fermion gas problems. In the latter, the reduction of spatial dimension
 is well known to enlarge the range of fluctuations which have a peculiar
impact on the properties of the system.\cite{Bychkov66,Dzyaloshinskii72} In this context the scaling
theory 
 commonly termed   multiplicative renormalization group,\cite{Menyhard73,Solyom79}  contributed
significantly to the completion of a coherent microscopic picture of the 1D fermion gas system as a 
paradigm of a non Fermi $-$ Luttinger $-$ liquid state. However,
this  scaling theory rests for the most part  upon the logarithmic structure of infrared singularities that
compose 1D perturbation theory and as such  departs somewhat from the
standard Wilson picture. It  remains closer to earlier  formulations of the
renormalization group in quantum field theory. Yet, fluctuations that characterize the 1D fermion gas do show
a multiplicity of length scales stretching from the shortest  distance, of the order of the lattice constant, 
up   to the quantum coherence length  or the de Broglie wavelength of fermions. Early attempts to reexamine
the properties of low dimensional  fermion gas system along these lines  were motivated by the  coupled
chains problem,\cite{Bourbon85a,Bourbon85,Bourbon86,Bourbon91,Bourbon95} which finds concrete applications
in the physics of quasi-one-dimensional organic conductors.
\cite{Jerome82,Bourbon99}  The description of  interacting fermions when  strong  spatial anisotropy is
present  shares many traits with its counterpart  in  critical phenomena.\cite{Grover73,Pfeuty74}  Hence the
Wilson method  provides a powerful framework to study how deviations  from perfect  1D scaling  are 
introduced by spatial anisotropy.   The  corresponding length scales enter as key components of the
notoriously complex description of dimensionality crossovers in which the 1D state evolves either towards the
emergence of a Fermi liquid or the formation of long-range order at finite temperature. 
 
 The  Wilson  renormalization group approach has attracted increasing interest in several areas of the
 many-fermion problem,\cite{Shankar94} notably in the description of  two-dimensional  systems in connection
with high-T$_c$ materials,\cite{Zanchi00,Halborth00,Honerkamp01} Fermi
liquids,\cite{Chitov95,Dupuis98,Dupuis96} 
 and ladder
systems.\cite{Fabrizio93,Lin97,Kishine98} Within the bounds of this review, it would  therefore  be
impossible to give anything more than a cursory account  of the expanding literature in
this field\cite{Zanchi01,Binz01,Salmhofer99}. With the goal of keeping  this paper  self-contained for the
non-specialist, however, our interest will be more selective and will focus on  the application of the 
renormalization group to the quasi-one-dimensional fermion gas  at finite temperature.\cite{notebose} 
Although many aspects of this problem have already been discussed in  detail in previous
reviews,\cite{Bourbon91,Bourbon95}  it is useful to revisit   some issues while in addition discussing
features of the Wilson method that deserve closer examination but which have  as yet never received a
detailed investigation.  This is especially true in regard  to the formulation of the  method itself, in
particular   the way  successive integrations over fermion  states in outer momentum shells (the Kadanoff
transformation) is carried out when  high order calculations are performed.  We close this review by a
detailed discussion of the dual nature of the Kohn-Luttinger mechanism\cite{Kohn65} from which interchain
pairing  correlations find their origin in  quasi-one-dimensional
systems.\cite{Emery86,Caron86,BealMonod86,Bourbon88} With the aid of the renormalization
group,\cite{Guay99,Guay97} we show how this mechanism  may yield instabilities of the metallic state towards
either unconventional superconducting or density-wave  order.

In section II, we introduce scaling notions for fermions in low dimension from a phenomenological
standpoint and predictions of the scaling hypothesis are compared to the solution in the non-interacting
case in section III. The formulation of the renormalization group in the classical Wilson scheme is given in
section IV and explicit calculations are carried out up to the two-loop level. In  section V, the influence
of interchain coupling is discussed and we go through various possibilities  of dimensionality
crossovers in the quasi-one-dimensional Fermi gas. In section VI, we study  how the Kohn-Luttinger mechanism
for interchain pairing correlation emerges from the renormalization group flow and under what conditions  it
can lead  to long-range order. We conclude in section VII.

\section{Scaling ansatz for fermions}
Before embarking on the program sketched above, it is first useful to discuss on purely phenomenological
grounds how  scaling notions apply to 1D  and  quasi-1D fermion systems.  The predictions of the
scaling  ansatz are  explicitly checked in the trivial but instructive case of the non-interacting Fermi
gas.   
\subsection{One dimension}
\label{Scaling1D}
Let us look at how   scaling ideas apply to the description of
purely one-dimensional fermionic systems at low temperature. The
temperature
$T$ is the parameter that controls the quantum delocalization or the spatial
coherence of each particle in
one dimension. This appears as a characteristic length
scale, 
$\xi$,  which is precisely the de Broglie wavelength.    The temperature dependence of $\xi$ 
is readily  determined by
considering the effect of thermal excitations on fermion states  within an
energy shell  $ \delta
\epsilon(k) \sim T$  ($k_B\equiv 1,\hbar\equiv 1$), where
$\epsilon(k)$ is the energy spectrum of fermions evaluated with respect to
the Fermi level. In the vicinity of
the Fermi points $\pm k_F$, one has $\delta
\epsilon(k)
\approx v_F (\mid k\mid - \ k_F)$, where $ v_F$ is  the Fermi velocity, and 
the coherence
length can be written 
\begin{equation}
\xi \sim v_F /T \sim T^{-\nu},
\label{coherence}
\end{equation}
where  the exponent $\nu$ is then equal to unity. As $T\to 0$,  $\xi $
goes to infinity and the
system develops long-range
quantum coherence.
    The `time' $\tau $ required for quantum
delocalization up to  $\xi$ is therefore  $\sim 1/T$  which simply implies that
\begin{equation}
\tau \sim \xi^z.
\end{equation}
where $z=1$ is the dynamical exponent of the fermion field.
Since `time' and distance must be considered on the same footing, both $\nu$ and $z$
can be taken  as independent of interaction.

Let us now consider the Matsubara-Fourier transform 
\begin{equation}
G(k,\omega_n) = \int_0^{\beta}\!\! \!\int_0^L e^{-ikx \,+\,
i\omega_n\tau}\,G(x,\tau)
\,  dxd\tau,
\end{equation}
where
\begin{equation}
G(x,\tau)= -\, \langle T_\tau
\,\psi_\alpha(x,\tau)\psi_\alpha^\dagger(0,0)\rangle
\end{equation}
is the one-particle time-ordered correlation function expressed as a statistical average
over fermion fields $\psi_\alpha^{(\dagger)}$ of
spin $\alpha$. Here
$\omega_n = (2n+1)\pi T$ 
 corresponds to the Matsubara frequencies. For non interacting fermions and $T=0$, $G(k) \sim (\mid k\mid- \
k_F)^{-1 }$  has a simple pole singularity at $|k|\to k_F$ and $\omega_n=0$, indicating that single particle
excitations are well defined. However, when fermions interact in one dimension, such a picture is expected
to be altered. In the scaling picture, this can be illustrated by adding an anomalous power dependence  on
the wave vector near $k_F$, namely 
\begin{equation}
    G(k) \sim (\mid k\mid- \ k_F)^{-1 +\theta}.
\label{scalingGk}
\end{equation}
  The absence of a quasi-particle pole introduces $\theta$, the
anomalous dimension for the  Green's function  ($
\theta\ge 0$ where the equality occurs in the non-interacting limit). Correspondingly, the absence of a Fermi
liquid component at zero temperature will affect the  equal-time decay of coherence at large distance, which
takes the form
\begin{equation}
G(x) \approx {\hat{C}\over x^{\bar{d} -1 +\theta} },
\label{Gvsx}
\end{equation}
  where the effective
dimension $\bar{d}= 2$ has a space and time component. A
similar decay in time takes place for
large
$\tau$ at $x=0$. Looking at the temperature dependence of the Green's function, the
scaling hypothesis  allows us to write
\begin{equation}
G(k,\omega_n) \approx CT^{-\bar{\!\gamma}}g\big(\delta k\,\xi,\omega_n\,
\xi^z\big),
\label{scalingG}
\end{equation} 
    where $\delta k= |k|-k_F$, $\bar{\!\gamma}$ is the thermal exponent of
the single particle Green's function,
and
$g(x,y)$ is a scaling function. Consistency between Eqns.~(\ref{scalingG}) and
(\ref{scalingGk}) leads to the following
relation between the exponents
\begin{equation}
\bar{\!\gamma}= (1-\theta)\nu.
\label{law1}
\end{equation}

Scaling can also be applied to the free energy per unit  length, the
temperature dependence of which can be
expressed as
\begin{eqnarray}
f = - {T\over L} \ln Z  \approx  A T^{1-\alpha}.
\end{eqnarray}
    The exponent $\alpha$ is connected to the temperature variation of
specific heat
\begin{eqnarray}
C_L =  && -T {\partial^2 f\over \partial T^2}\cr
     \approx && A' T^{-\alpha},
\end{eqnarray}
where
\begin{eqnarray}
\alpha= 1-\bar{d}\nu.
\end{eqnarray}
As long as the
ansatz $\nu=1$ holds for the
coherence length, one then always has $\alpha=-1$ in                    one
spatial
dimension. The specific heat is therefore
linear  in temperature for fermions even though, according
to  Eqn. (\ref{scalingGk}), the system is no longer a Fermi liquid  at low
energy. It should be noted
that the familiar Sommerfeld  result $ C_V \sim T$ for a Fermi liquid
(and Fermi gas)
in higher dimensions  indicates that
the importance put on fermion states near the Fermi level
one-dimensionalizes
the
sum over states for the evaluation of internal energy.

Scaling concepts can also be applied to the critical behavior of
correlations that involve {\it pairs} of
fermions in one dimension.  Superconducting and  $2k_F$ density-wave
responses are among the pair
susceptibilities  that may present singularities at zero temperature. Consider
the Matsubara-Fourier transform of the susceptibility
\begin{equation}
\chi(q,\omega_m) = \int_0^{\beta}\!\! \!\int_0^L    \, e^{-iqx
+i\omega_m\tau}\chi
(x,\tau) \, dx d\tau ,
\end{equation}
    where  $\omega_m=2m\pi T$. The time-ordered pair correlation function is
\begin{equation}
\chi(x,\tau) = - \langle T_\tau O(x,\tau) O^\dagger(0,0)\rangle,
\label{defchi}
\end{equation}
    in  which $O^\dagger\sim
\psi^\dagger_\alpha\psi^\dagger_\beta$
($O^\dagger\sim \psi_\alpha^\dagger\psi_\beta$) are
    the composite operators of  the superconducting (density-wave) channel, commonly called the 
{Cooper}
({Peierls}) channel.
    At  $T=0$, these can show long-range coherence  leading
to an algebraic decay of pair correlations
with  distance. Following the example of the correlation of the order parameter at a critical
point,  one can write
   a relation of the form
\begin{equation}
\chi(x) \approx  e^{iq_0x}{D\over  x^{\bar{d}-2 +\eta}}~,
\label{chix}
\end{equation}
where $\eta\ge 0 $ is an anomalous exponent that characterizes the decay of pair
correlation. Here $q_0=0 \ (2k_F)$ is the characteristic wave vector of correlations in the Cooper (Peierls)
channel. A similar power law variation with time
$\tau$ can also be written
at small distance. Correspondingly, in Fourier space, one has in the static
limit
\begin{equation}
\chi(q) \approx  {\bar{D}\over q^{2-\eta}}~,
\label{chiq}
\end{equation}
    where $q$ refers to  deviations with respect to an ordering wave vector $q_0$ in the Cooper or Peierls
channel. For
$q=0$ and at finite frequency, a similar expression holds with
$q$ being replaced by the frequency. Now at finite temperature, the scaling
hypothesis allows us to write
\begin{equation}
\chi(x, T) \approx D  e^{iq_0x} x^{- \bar{d}+2 -\eta} {\cal X}(x/\xi),
\end{equation}
where ${\cal X}$ is scaling function (${\cal X}(0)=1$ ). Here the
correlation length $\xi$ for the
pair field is assumed to have the same  power law dependence on temperature as the
coherence length Eqn.~(\ref{coherence})
for  both fermions in  a pair. In Fourier space,  one can write for the
divergence of the susceptibility in a
given channel 
\begin{equation}
\chi(q,\omega_m,T) \approx \bar{D} T^{-\gamma} \bar{\cal
X}(q\xi,\omega_m\xi^z).
\label{chiqs}
\end{equation}
    Consistency
between  Eqns.~(\ref{chiqs}) and (\ref{chiq}) in the static case leads to the
Fisher relation
\begin{equation}
\gamma= (2-\eta)\nu
\label{scalinglaw}
\end{equation}
between the exponents of the  static susceptibility, the correlation
function and the correlation length.

\subsection{Anisotropic scaling and  crossover phenomena}

Let us  now move from one-dimensional systems to quasi-one- dimensional systems by which we mean weakly
coupled chains. A dimensionality crossover  in low dimensional  fermion systems is induced
by a small coupling between chains. The nature of this crossover will depend on the kind of   coupling
involved.  Weakly coupled chains correspond to
      so-called quasi-one-dimensional anisotropy, a situation realized in
practice for
electronic
materials such as the  organic  conductors,\cite{Jerome82,Bourbon99,Ishiguro90} lattice of spin
chains, etc. In ordinary critical phenomena, scaling concepts were extended to anisotropic  systems and have
been successful in  describing the new  length scales introduced by anisotropy;  
\cite{Riedel69,Pfeuty74,Grover73} their use  in the  study  of
   anisotropic fermion systems appears therefore quite natural.  For
example, deconfinement of
 quantum coherence for single particles  can give rise to the restoration of a
Fermi liquid component below some
characteristic low energy scale. In addition, the deconfinement of
pair correlations in more than
one spatial dimension can lead to the emergence of  true long-range order
at finite temperature.\cite{Mermin66}

\begin{figure} 
\centerline{\includegraphics[width=7cm]{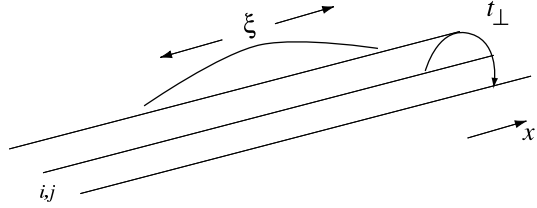}}
\caption{One-fermion interchain hopping.   $\xi$ is the fermion coherence length along the chains. 
}\label{hopping}\end{figure}

Let us consider  first the case of interchain
single-particle  hopping (Fig.~\ref{hopping}), whose
amplitude $t_\perp$ is small compared to the Fermi energy
of isolated
chains ($t_\perp \ll E_F$). Applying the extended scaling ansatz to the
single fermion Green's
function at $\mid k\mid= k_F$, and $\omega_{n=0}=\pi T$,\cite{Pfeuty74} one can write
\begin{equation}
G(T,k_\perp)\approx  DT^{-\bar{\gamma}}X_G(Bt_\perp(k_\perp)/T^{\phi_x}).
\label{crossX}
\end{equation}
     $t_\perp(k_\perp)= t_\perp \cos k_\perp $ and $k_\perp$
is the transverse momentum of the
Fermion spectrum (here the inter-chain distance $d_\perp$ has been set  equal to unity).    $\phi_x>0$
is the crossover exponent which
governs  the magnification of the single-particle perturbation
$t_\perp$
     as $\xi$ increases along the chains (Fig.~\ref{hopping}).
$X_G(y)$ is called a crossover scaling function and it determines
transitory aspects of the crossover
and the  temperature range over which it is achieved. Here $X_G(0) =1$.
When $y\ll1$, the
coherence of the fermionic  system is   confined along the chains
and  the physics is
essentially dominated by the one-dimensional expression (\ref{scalingG}).
Otherwise for $y \gg 1$, a different 
temperature dependence  is expected to arise. The
temperature scale for the
single-particle crossover is thus determined by the condition $y\sim 1$,
that  is
\begin{equation}
T_x\sim t_\perp^{1/\phi_x}.
\end{equation}
Well below  $T_x$  and on the full Fermi surface ${\bf k}={\bf k}_F$, one has
\begin{equation}
G(T) \sim z(t_\perp) T^{-\dot{\bar{\gamma}}},
\label{crossXb}
\end{equation}
where $z(t_\perp)$ is the quasi-particle weight and $\dot{\bar{\gamma}}=1$ is the Fermi liquid exponent
that restores the canonical dimension of the
Green's  function. In order to match (\ref{crossX})
and (\ref{crossXb}) well below $T_x$, one
must have \cite{Pfeuty74}
\begin{equation}
z(t_\perp)= z_\infty \ t_\perp^{-(\bar{\gamma}-\,\dot{\bar{\gamma}})/\phi_x},
\label{coefscale}
\end{equation}
    where $z_\infty$ is a non universal constant. Applying scaling
    to the dynamical $-$ coherent $-$  part of the single-particle Green's
function,
the recovery of 
Fermi liquid behavior yields
\begin{equation}
G({\bf k},\omega_n) \approx {z(t_\perp)\over [i\omega_n -E({\bf
k})]^{\dot{\bar{\gamma}}}} \
\label{scalingG3D}
\end{equation}
with $ \dot{\bar{\gamma}}=1 $ again.

 Interchain coupling will  affect the temperature dependence of pair
correlations as well,
which can undergo a dimensionality crossover and in turn an instability
towards the formation of
    long-range order  either in the Cooper or the Peierls channel at finite
temperature. This
  possibility is of course bound to the number of components of the
order parameter and the spatial
dimensionality of the system.\cite{Mermin66} The scaling behavior of pair
correlations is therefore analogous to the one found in ordinary
anisotropic critical phenomena. Properties of
the system   close
to the true critical point at $T_c$ can then be described by  the usual reduced
temperature  
\begin{equation}
\dot{t} = {T- T_c\over T_c}.
\end{equation}
The static pair susceptibility  at some ordering wave vector ${\bf
Q}_0\equiv(q_0,q_{\perp0})$ will then diverge as $\dot{t}\to 0$
according to
\begin{equation}
\chi(T) \approx \dot{C}(g_\perp)\,\dot{t}^{-\dot{\!\gamma}},
\label{chitc}
\end{equation}
where $\,\dot{\!\gamma}$ is the susceptibility exponent.

Applying the extended scaling hypothesis to the static susceptibility in the
superconducting or
the density-wave  channel, one can
write
\begin{equation}
\chi(T) \approx \bar{D}T^{-\gamma} X ( B g_{\perp }/T^{\phi_{x^2}}),
\label{chiX}
\end{equation}
where $X(y)$ is a crossover scaling function $(X(0)=1)$, $g_{\perp }$ is
the inter-chain
coupling for pairs of fermions and $\phi_{x^2}$ is a {\it two}-particle crossover
exponent.\cite{Bourbon91,Bourbon93}
Following the same arguments as  in the  single-particle case, the change in the
exponents will occur
when $y\sim 1$, which allows us to introduce the crossover temperature for pair correlations
\begin{equation}
T_{x^2} \sim g_\perp^{1/\phi_{x^2}}.
\label{cross2}
\end{equation}
It is worth noting that the interchain coupling  $g_\perp$  for pairs of
particles differs from $t_\perp$,
   and  the two-particle crossover can occur above its single particle counterpart
$T_x$, in which case the restoration of  Fermi liquid behavior is absent. Instead  only pair
correlations  deconfine, possibly leading  to long-range order.

If the  scaling hypothesis holds, the expressions
(\ref{chitc}) and (\ref{chix}) have to
match near $T_c$, which implies
\begin{equation}
\dot{C}(g_\perp) = C_\infty \ g_\perp^{-(\gamma -\dot{\gamma})/\phi_{x^2}},
\label{constraint}
\end{equation}
where $C_\infty$ is  a non universal constant. Another prediction of the scaling hypothesis concerns the
shift in
$T_c$ from zero temperature\cite{Pfeuty74}
\begin{equation}
T_c(g_\perp) \sim T_x \sim g_\perp^{1/\phi_{x^2}}
\label{scalingTc}
\end{equation}
When the system undergoes its two-particle dimensionality crossover, the
temperature dependence of
singularities near $T_c$ is governed by the correlation length
\begin{equation}
\dot{\vec{\!\xi}}= \,\dot{\vec{\!\xi}}_0\ \dot{\!t}^{-\dot{\nu}},
\label{correlation}
\end{equation}
    where \ $\dot{\vec{\!\xi}}_0$ is the coherence length at $T\sim
T_x$, which  fixes a
short anisotropic length scales for the
onset of critical fluctuations described by the exponent $\dot{\nu}$.
   At $T_c$, one has the critical form
\begin{equation}
\chi({\bf q}) \approx {\dot{D}\over \big(\dot{\xi}^2_{0\parallel}q^2 +
\dot{\xi}^2_{0\perp}q_\perp^2\big)^{\dm(2-\dot{\eta})}},
\end{equation}
which according to (\ref{chitc}) leads to the Fisher relation
\begin{equation}
\dot{\gamma}=(2-\dot{\eta})\dot{\nu}.
\end{equation}

\section{Free fermion limit}
Before facing the task of calculating the exponents and  various scaling
quantities of interacting fermion models,
it is  instructive to   first analyze the properties of the free fermion gas
model. We will consider first the purely 1D system and then introduces interchain single particle
hopping  and see how they actually satisfy
    the above scaling relations.
\subsection{One dimension}
    Let us consider  a
one-dimensional
system of free fermions coupled to source pair  fields $h$ for which the
partition
function is given by   a functional
integral  over the anti-commuting (Grassman) fields $\psi$ (see  references~\cite{Negele88,Shankar94} for
reviews on Grassman fields)
\begin{eqnarray}
Z_0[h^*,h] =  && \int\!\!\int {\cal D} \psi^*{\cal D}\psi \
e^{S_0[\psi^*,\psi] +S_h[\psi^*,\psi,h^*,h]}\cr
            = && \int \!\!\int {\cal D} \psi^*{\cal D}\psi \
\exp\Big\{\sum_{\kktil,\alpha}\,
[G_p^0(\kktil)]^{-1}\psi_{p,\alpha}^*(\kktil) \psi_{p,\alpha}(\kktil)
+ \sum_{\mu,\qtil} \, [
O_\mu^*(\qtil)h_\mu(\qtil) + {\rm c.c}\ ]
\Big\}.
\end{eqnarray}
    Here $S_0$
is the
action  of the system in the absence of fields and 
\begin{eqnarray}
G_p^0(\kktil)= && -\langle\psi(\kktil)\psi^*(\kktil)\rangle_0\cr
                = && [i\omega_n - \epsilon_p(k)]^{-1} 
\end{eqnarray}
is the free fermion propagator in the  Fourier-Matsubara space where
$\kktil=(k,\omega_n)$. The fermion spectrum
is linearized with  respect to the  Fermi points $\pm k_F$, namely
\begin{equation}
\epsilon_p(k)= v_F(pk-k_F),
\end{equation}
    where $p=\pm$ refers to right $(+)$
and left $(-)$ going fermions and $v_F$ is the Fermi velocity.
A natural cutoff on the bandwidth can be imposed on this continuum  model
by restricting the
summations on the wave vector $k$ of 
branch  $p$ in  the interval $pk_F + k_0 > k > pk_F-k_0$.  The cutoff $k_0\sim 1/a$,
where
$a$ is the lattice constant
    on the band wave vector,  then leads to a finite bandwidth $E_0= 2v_Fk_0
\equiv 2E_F  $ which we will take
equal to twice the Fermi energy $E_F=v_Fk_F$.

\begin{figure}[htb] \centerline{\includegraphics[width=7cm]{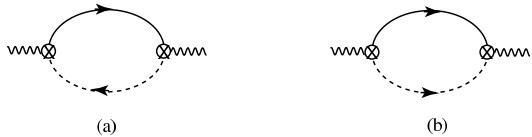}}
\caption{Diagrams for the bare susceptibilities in the Peierls (a) and
Cooper (b)
channels. A continuous (dashed) line correspond to a particle or hole near
$+k_F$
($-k_F$). }\label{Bulles}\end{figure}

In one dimension,
    the fermion gas is unstable to
the formation of  pair correlations   in both   Cooper and
Peierls channels. This shows up as infrared
singularities for  pair
susceptibilities, which can be calculated by adding  $S_h$ to the action, which linearly couples  source
fields to fermion pairs.
In the Peierls channel ($\mu=\mu_P$), the particle-hole fields
\begin{equation}
O_{\mu_P=0}(\qtil) = \sqrt{T\over L}\sum_{\kktil,\alpha}
\psi^*_{-,\alpha}(\kktil-\qtil)\psi_{+,\alpha}(\kktil)
\end{equation}
and
\begin{equation}
O_{\mu_P\ne 0}(\qtil) = \sqrt{T\over L}\sum_{\kktil,\alpha,\beta}
\psi^*_{-,\alpha}(\kktil-\qtil)\sigma_{\mu_P}^{\alpha\beta}\psi_{+,\beta}(\kktil)
\end{equation}
correspond to charge-density-wave (CDW: $\mu_P=0$) and
spin-density-wave (SDW:$\mu_P\ne0$)
correlations respectively near the wave vector $2k_F$, where $\qtil=
(2k_F +q, \omega_m)$ ,
$\sigma_{\mu_P=1,2,3}$ are the  Pauli matrices and $L$ is the length of the
system. In the
Cooper channel ($\mu=\mu_C$), the particle-particle fields
    \begin{equation}
{ O}_{\mu_C=0}(\qtil) = \sqrt{T\over L}\sum_{\kktil,\alpha}
\alpha \ \!\psi_{-,-\alpha}(-\kktil+\qtil)\psi_{+,\alpha}(\kktil)
\label{singulet}
\end{equation}
and
\begin{equation}
{ O}_{\mu_C\ne 0}(\qtil) = \sqrt{T\over L}\sum_{\kktil,\alpha,\beta}
\psi_{-,-\alpha}(-\kktil+\qtil)\sigma_{\mu}^{\alpha\beta}\psi_{+,\beta}(\kktil),
\label{triplet}
\end{equation}
correspond to singlet  (SS: $\mu_C=0$) and triplet (TS: $\mu_C\ne0$,)
superconducting channels
respectively, where $\qtil=(q,\omega_m)$.
The  susceptibilities in zero $h_{\mu}$ in both channels
are defined by
\begin{eqnarray}
\chi_\mu(\qtil) = &&- {1\over Z_0} {\delta^2 Z_0[h^*,h]\over \delta
h_\mu^*(\qtil)\delta h_\mu(\qtil)}\Big\vert_{h=0}\cr
= && - \,\langle O_\mu(\qtil)O^*_\mu(\qtil)\rangle_0,
\end{eqnarray}
    and  obtained  from the calculation of the
statistical averages
\begin{equation}
\chi_\mu(\qtil) = -\ Z_0^{-1} \int\!\!\int {\cal D} \psi^*{\cal D}\psi \
O_\mu(\qtil)O^*_\mu(\qtil)\ e^{S_0[\psi^*,\psi]}.
\end{equation}
in the Peierls and Cooper
channels. Thus in the Peierls channel, the dynamic  susceptibility
(Figure\ \ref{Bulles}-a)
close to $2k_F$  is given by the standard result
\begin{eqnarray}
\chi_{\mu_P}(\qtil)&= &  2{T\over L} \sum_{\kktil}
G_-^0(k-2k_F-q,\omega_n-\omega_m)G_+^0(k,\omega_n)\cr
    &= & - N(0)\Big\{ \ln {1.13 E_F\over T} + \, \psi(1/2) -
\dm\Big[\psi\Big(\, 1/2 + {iv_Fq -\omega_m\over 4\pi
T}\,\Big) + {\rm c.c }\, \Big]\Big\}
\label{chiPeierls}
\end{eqnarray}
for all density-wave components $\mu$  at low
temperature. Here  $N(0)=1/(\pi v_F)$ is the
density of states per spin at the Fermi level and $\psi(x)$ is the digamma function. In the evaluation of
the Peierls bubble, the electron-hole symmetry
\begin{equation}
\epsilon_+(k) = - \epsilon_-(k-2k_F),
\end{equation}
of the  fermion spectrum $-$ called nesting $-$ has been  used in the  evaluation of the Peierls bubble. The
elementary  susceptibility in the Cooper channel  (Figure\ \ref{Bulles}-b) has a similar behavior, that is
\begin{eqnarray}
\chi_{\mu_C}(\qtil)&= &  - 2{T\over L} \sum_{\kktil}
G_-^0(-k+q,-\omega_n+\omega_m)G_+^0(k,\omega_n)\cr
 &  = &  - N(0)\Big\{ \ln {1.13 E_F\over T} + \, \psi(1/2) -
\dm\Big[\psi\Big(\, 1/2\, + {iv_Fq -\omega_m\over 4\pi
T}\,\Big)\, + \,{\rm c.c} \,\Big]\Big\}
\label{chiCooper}
\end{eqnarray}
for which the inversion symmetry (which is independent of the spatial dimension)
\begin{equation}
\epsilon_+(k) =\epsilon_-(-k)
\label{Cooper}
\end{equation}
of the spectrum has been
used. 

\begin{figure}[htb]\centerline{\includegraphics[width=0.7\hsize]{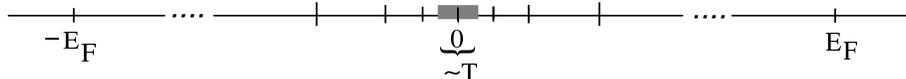}}
\caption{Energy scales for the 1D fermion problem. Between the Fermi ($E_F$) and thermal energy ($T$),
there is no characteristic energy scale.}\label{Scale}\end{figure}

In the static limit, both channels show  logarithmic singularities of the
form $\chi_\mu(T)\sim \ln(E_F/T)$
indicating that pair correlations have no
characteristic energy scale between  $E_F$ and $T$. All
intermediate scales between $E_F$ and $T$  turn out to give similar
contributions to the integrals (\ref{chiPeierls}) and 
(\ref{chiCooper}), which are roughly of the 
form $\int_T^{E_F} d\epsilon_+/\epsilon_+$ at $q=0$ and $\omega_m=0$ as shown in Fig.~\ref{Scale}, a feature
that can be equated with scaling.\cite{Wilson75}

This logarithmic divergence can also be seen as a power law with a vanishing
exponent, that is
\begin{equation}
\chi_\mu(T) = \lim_{\gamma\rightarrow 0^+}-N(0)
\gamma^{-1}\Big[\Big({T\over 1.13 E_F}\Big)^{-\gamma} -1\Big].
\end{equation}
   According to the relation  (\ref{scalinglaw}), this should imply $\eta\to
2$, namely a
logarithmic singularity in  $q$. This is indeed consistent with the
asymptotic behavior of $ \chi_\mu(q)$  at
$T=0$ and  $\omega_m=0$, which is found to be
\begin{equation}
\chi_\mu(q)= N(0)\ln {|v_Fq|\over 2 E_F},
\label{asymptoq}
\end{equation}
 in agreement with (\ref{chiq}).

The origin of scale invariance can be further sharpened if one looks  at
the decay of the fermion
coherence at large distance. Consider the Green's function
\begin{eqnarray}
G_p^0(x,\tau) & =  & {T\over L} \sum_{k,\omega_n} G_p^0(k,\omega)\  e^{ikx
-i\omega_n\tau}\cr
                 & = &  {e^{ipk_Fx}\over2\pi  i} {1\over \xi\ {\sinh
[(x+ipv_F\tau)/\xi]} }
                 \end{eqnarray}
where
\begin{equation}
\xi = {v_F\over \pi T}
\end{equation}
   is a characteristic length scale  for  the
quantum coherence of each fermion and in fact the de Broglie
quantum length.  At low temperature for
$a \ll x\ll \xi$, one finds the power law decay of coherence  as a function
of distance
\begin{equation}
G^0_p(x) \sim
1/x,
\end{equation}
which signals homogeneity or scaling. Reverting to (\ref{Gvsx}), we verify that $\theta= 0$  in the free
fermion limit.
    For $ x\gg \xi$,
\begin{equation}
G^0_p(x)\sim e^{-x/\xi}
\end{equation}
   there is no homogeneity and the coherence is  exponentially  damped  with
distance.   The absence of
energy scale between  $E_F$ and $T$  thus corresponds in real space to
   the absence of length scale between $\xi$ and $a$.
Rewriting the elementary   the Cooper and Peierls
susceptibilities  as
\begin{equation}
\chi_\mu(T) \approx - N(0)\ln {\xi\over a},
\end{equation}
one then  sees that the quantum spatial coherence of particles becomes   an essential
ingredient  in building up
logarithmic singularities in  pair correlations of the Cooper and Peierls
channels.
  This can be directly verified by 
the decay of pair correlations in space and time. From the definition (\ref{defchi}) and the use of Wick
theorem for free fermions, one obtains   
\begin{eqnarray}
\chi_\mu(x,\tau)&= & \mp 2\, G^0_-(\mp x,\mp\tau) G^0_+(-x,-\tau)\cr
                    &=  &   e^{-iq_0x}{1\over 2\pi^2\xi^2 } {1\over
{\sinh[(x+iv_F\tau)/\xi]}\  {\sinh}[(x-iv_F\tau)/\xi]},
\end{eqnarray}
where the upper (lower) sign corresponds to the Cooper (Peierls) case. For $  a\ll x \ll \xi $, one finds
\begin{equation}
\chi_\mu(x) \propto  { e^{iq_0x}\over x^2},
\end{equation}
   in agreement with (\ref{chix}), in which $\bar{d}=2$
and $\eta=2$. At large distance,
when $x\gg \xi$, one has
\begin{equation}
\chi_\mu(x) \propto  e^{iq_0x}  e^{-2x/\xi},
\end{equation}
where the effective coherence length for  pairs is half the one of a single
fermion.
\subsection{Interchain coupling}
\label{sechopping}
A very simple illustration of a dimensionality crossover for the
single-particle coherence can be readily
given  in the free fermion case by adding  an interchain  hopping term to
the  the free action, which becomes
\begin{eqnarray}
S_0[\psi^*,\psi] &= & S_0[\psi^*,\psi]_{1D}\  +
\sum_{\kktil,p,\alpha}\sum_{\langle i,j\rangle} t_{\perp}
\psi_{p,\alpha,i}^*(\kktil)\psi_{p,\alpha,j}(\kktil)\cr
&= & \sum_{\kvec,\omega_n}\sum_{p,\alpha} [G^0_p(\kvec,\omega_n)]^{-1}
\psi_{p,\alpha}^*(\kvec,\omega_n)\psi_{p,\alpha}(\kvec,\omega_n)
\end{eqnarray}
where $t_{\perp}$ is the single-particle hopping integral between
nearest-neighbor chains $i$ and  $j$.
Considering  a linear array of $N_\perp$ chains, the free propagator that
parameterizes $S_0$
in Fourier  is 
\begin{equation}
G^0_p(\kvec,\omega_n) = \big( i\omega_n - E_p(\kvec)\big)^{-1},
\label{tperp}
\end{equation}
where 
\begin{equation}
E_p(\kvec) = \epsilon_p(k) - 2t_\perp \cos k_\perp
\end{equation}
is the full fermion spectrum. The propagator
can be trivially  written in the extended scaling scaling form (\ref{crossX}) with
$\bar{\gamma}=1$, where the scaling function
takes the simple form
\begin{equation}
X_G(y)= (1 + y)^{-1},
\end{equation}
with $y= Bt_\perp\cos(k_\perp)/T$ and $B=(i\pi/2)^{-1}$. This
implies $\phi_{x^1}=1$ for the
single-particle crossover exponent. This value of $\phi_{x^1}$ is found
in  the anisotropic free field
theory of critical phenomena.\cite{Pfeuty74}   The Fourier transform leads
to single-particle fermion
correlation function
\begin{equation}
G^0_p(x,n_\perp,\tau)= G^0_p(x,\tau) e^{-i\pi n_\perp}J_{n_\perp}(x/\xi_x),
\end{equation}
where $n_\perp$ is the transverse separation expressed in number of chains. The
crossover scaling function is then  embodied in the Bessel function
$J_{n_\perp}(x/\xi_x)$   of  order
$n_\perp$, which takes sizable values when the intrachain distance $x$
reaches values of the
order of the characteristic length scale $\xi_x = v_F/(2t_\perp)$ needed for
coherent hopping from one chain to
another (Figure~\ref{hopping}). For  $x\sim \xi$, the crossover condition is $\xi \sim \xi_x $,
 and this leads once again to $T_{x^1} \sim
t_\perp$, namely $\phi_{x^1}=1$.

%
\section{The Kadanoff-Wilson renormalization group}
\subsection{One-dimensional case}
\label{KW}
\begin{figure}[htb] \centerline{\includegraphics[width=8cm]{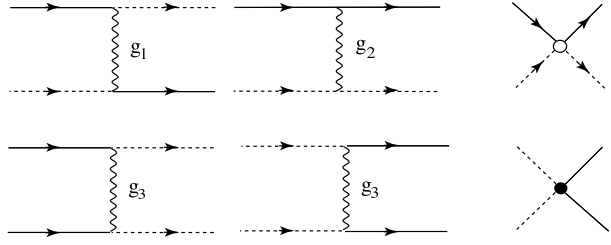}}
\caption{Backward ($g_1$), forward ($g_2$) and umklapp ($g_3$) couplings of
the 1-D fermion gas model and the
corresponding diagrams. The open (full) circle  corresponds to the generic
vertex part for backward and forward
(umklapp) scatterings. }\label{Intg1g2}\end{figure}

When the  motion  of fermions is entirely confined to one spatial dimension,
particles  cannot avoid each other  and hence the influence of  interactions is particularly important. The
absence of  length scale up to $\xi$ found in the free
fermion gas problem for both single and pair correlations  will carry over
in the presence of interactions, which  couple  different energy scales. 
The  low energy behavior of the
system  in which we are interested  will be
influenced by all the higher energy states up to the
highest energy at the band edge. Renormalization group ideas can be used
to tackle the problem of
the cascade of energy or length scales that characterizes interacting fermions. This will be
done in the framework of the fermion gas model in which 
 the direct interaction between fermions is parametrized  
by a small set of distinct  scattering processes close to the Fermi level (Figure \ref{Intg1g2}).
\cite{Dzyaloshinskii72}  This is justified  given   the particular importance shown  by the Cooper and
Peierls infrared singularities for electronic states near the Fermi level. We will consider the
backward ($g_1$) and forward  ($g_2$) scattering processes for which particles on opposite sides of the Fermi
level exchange momentum near $2k_F$ and zero respectively. When the band is half-filled, $4k_F$ coincides
with a reciprocal lattice vector, making possible   Umklapp processes, denoted by 
$g_3$, in which  two particles  can be  transferred  from one side of the Fermi level to the
other.
      Thus,  adding the interaction part
$S_I$ to the action, the partition function in the absence of   source fields
takes the form 
\begin{eqnarray}
Z &= & \int\!\!\!\int {\cal D} \psi^*{\cal D}\psi \ e^{S_0[\psi^*,\psi] +
S_I[\psi^*,\psi]}\cr
           &= &\int\!\!\!\int {\cal D} \psi^*{\cal D}\psi \
\exp\Big\{\sum_{\kktil,\alpha}
[G_p^0(\kktil)]^{-1}\psi_{p,\alpha}^*(\kktil) \psi_{p,\alpha}(\kktil) \cr
& + &\  \pi v_FT/L \sum_{\lbrace
p,\kktil,\qtil,\alpha \rbrace} g_{\{\alpha\}}
   \ \psi^{\ast}_{+,\alpha_1}(\kktil_1+\qtil)\psi^{\ast}_{-,\alpha_2}(\kktil_2
-\qtil) \psi_{-,\alpha_3}(\kktil_2)\psi_{+,\alpha_4}(\kktil_1)\cr
&
-&\ \pi v_F\ T/2L \sum_{\lbrace p,\kktil,\qtil,\alpha \rbrace} g_3 \
\psi^{\ast}_{p,\alpha}(\kktil_1+p\qtil)\psi^{\ast}_{-p,\alpha'}(\kktil_1
-p\qtil + p\tilde{G})
    \psi_{-p,\alpha'}(\kktil_2) \psi_{p,\alpha}(\kktil_1)  \Big\}
\label{partition}
\end{eqnarray}
where
\begin{equation}
g_{\{\alpha\}} =
g_1\delta_{\alpha_1\alpha_3}\delta_{\alpha_2\alpha_4}  -
g_2\delta_{\alpha_1\alpha_4}\delta_{\alpha_2\alpha_3}
\label{gincom}
\end{equation}
and $\tilde{G}=(4k_F,0)$ is the reciprocal lattice vector. The parameter space of the action is then 
\begin{equation}
\mu_S = (G^0_p,g_1,g_2,g_3).
\end{equation}
where all the couplings are expressed in units of $\pi v_F$.

The idea behind the Kadanoff-Wilson renormalization group  is  the
transformation or renormalization of the action $S$ following  successive
partial integration of
$\bar{\psi}'$s of momentum located in the energy shell
   $\frac{1}{2} E_0(\ell)d\ell$ on both sides of the Fermi level, where
$E_0(\ell)=E_0e^{-\ell}$ is the effective bandwidth at step $\ell$ of
integration and $d\ell
\ll 1$. The transformation of $S$ from $\ell$ to $\ell +d\ell $,  that keeps
the partition
function invariant is usually written as
\begin{eqnarray}
Z=  && e^{A(\ell)} \int\!\!\!\int_< {\cal D}\psi^*{\cal D}\psi
e^{S[\psi^*,\psi]_<}\ \int\!\!\!\int{\cal
D}\bar{\psi}^*{\cal D}\bar{\psi}
\,e^{S[\psi^*,\psi,\bar{\psi}^*,\bar{\psi}]_{d\ell}} \cr
   = &&  e^{A(\ell+d\ell)}\int\!\!\!\int_< {\cal D}\psi^*{\cal D}\psi
e^{S[\psi^*,\psi]_{\ell +d\ell}},
\label{kadanoff}
\end{eqnarray}
where $A(\ell)$ corresponds to the free energy density at the step $\ell$. The  integration measure  in
the outer  shell is
\begin{equation}
{\cal D}\bar{\psi}^*{\cal D}\bar{\psi} =  \prod_{ p,\alpha,\{\kktil\}'
}
d\bar{\psi}^*_{p,\alpha}(\kktil)d\bar{\psi}_{p,\alpha}(\kktil).
\label{measure}
\end{equation}
and $\{\kktil\}'= \{k\}'\{\omega_n\}$. Here  $\{k\}'$ corresponds to the
momentum outer shells
$$  \ k_0e^{-\ell}+ k_F > k  > k_0e^{-\ell
-d\ell}+ k_F \ \ \ \ \ \ \ \  \ \ (-k_0e^{-\ell -d\ell} - k_F > k  > -k_0e^{-\ell
}- k_F)
$$
above the Fermi level and
$$
 -k_0e^{-\ell-d\ell}+ k_F> k  > -k_0e^{-\ell }+ k_F \ \ \ \ \ (-k_0e^{-\ell} - k_F > k  > -k_0e^{-\ell-d\ell}- k_F)
$$
   below for right (left) moving fermions, while $\{\omega_n\}$ covers all the Matsubara frequencies. The
remaining inner shell ($<$) fermion
degrees of freedom are kept fixed. The parameters of the action denoted as
$\mu_S$  
transform according to
\begin{equation}
R_{d\ell}\, \mu_S(\ell)= \mu_S(\ell+d\ell).
\end{equation}
 where 
\begin{equation}
\mu_S(\ell)= \big(z(\ell)G_p^0, z_1(\ell)g_1,z_2(\ell)g_2,z_3(\ell)g_3)
\end{equation}
contains the  factors $z$ and $z_{i=1,2,3}$ that renormalize the parameter space of the action at
$\ell$.\cite{note1}    The flow of parameters is conducted down to
$\ell_T\equiv
\ln E_F/T$  corresponding to the highest value of $\ell$ at which scaling
applies  as discussed in \S~II.  This leads in turn to the temperature dependence of the parameter space
$\mu_S(T)$. It should be mentioned that the Kadanoff transformation will  generate new terms which were not
present in the action at
$\ell=0$. In principle, the
relevance of these can be seen explicitly  from a rescaling of energy 
and  field `density' after each partial integration. Thus performing an energy change by the factor
$s=e^{d\ell}>1 $, one has
\begin{eqnarray}
\epsilon_p'= && s \epsilon_p\cr
\omega_n'= &&  s\omega_n. 
\label{rescaleE}
\end{eqnarray}
The fermion fields in the action will then transform according to 
\begin{equation}
\psi'^{(*)} = z^{-\dm} s^{-\dm}\psi^{(*)},
\label{rescaleF}
\end{equation}
which in its turn implies that the rescaling of interaction parameters is given by
\begin{equation}
g_{i}'= g_i\, z_i\,z^2, \ \ \ \ \  {\rm for }\ \ \ \  { i=1,2,3}.
\end{equation}
This  contains only   anomalous renormalization factors indicating that the interaction parameters are
all marginal. However, if interactions between three, four  or more particles   were added to the
{\it bare} action, the amplitude of these  would decay as $1/s$ to some positive  power, and
would therefore be irrelevant in the RG sense. A different conclusion would be reached, however, if  
these many-body interactions were generated in the flow of renormalization. In this case, the amplitudes of
interactions becomes scale dependent and such couplings can remain marginal or even become relevant (see
\S~\ref{Stwoloop} and \S~\ref{Spairhopping}).    Besides this use,   rescaling for our fermion problem is
not an essential step of the RG transformation. We will consider the renormalization of
$\mu_S$ with respect to 
$\ell$ as a transformation that describes an effective system with a reduced band width $E_0(\ell)$ or
 a magnified length scale 
$ v_F/E_0(\ell)$.

In practice, the  partial integration (\ref{kadanoff}) is most easily performed  with the
aid of diagrams. We first decompose the ${\kktil}$  sums in the
action into outer and inner shells momentum variables
\begin{equation}
\sum_{\{\kktil\}} = \sum_{\{\kktil\}'} + \sum_{\{\kktil\}_<}.
\end{equation}
 This allows us  to write
\begin{eqnarray}
S[\psi^*,\psi] = S[\psi^*,\psi]_< + S[\psi^*,\psi,\bar{\psi}^*,\bar{\psi}]
\label{action}
\end{eqnarray}
where $S[\psi^*,\psi]_<$ of the action with all the $\psi'$s in the inner
shell, whereas
\begin{eqnarray}
S[\psi^*,\psi,\bar{\psi}^*,\bar{\psi}]= S_0[\bar{\psi}^*,\bar{\psi}] +
\sum_{i=1}^4S_{I,i}[\psi^*,\psi,\bar{\psi}^*,\bar{\psi}]
\label{Sdecompos}
\end{eqnarray}
consists in a free part in the outer momentum shell and  an interacting
part as a sum of terms, $S_{I,i}$, 
having
$i= 1,\ldots 4 $ $\bar{\psi}'$s in the outer momentum shell. Their
integration following the partial trace
operation (\ref{kadanoff}) is performed perturbatively with respect to
$S_0[\bar{\psi}^*,\bar{\psi}]$.
Making use of the linked cluster theorem,  the outer shell integration becomes
\begin{eqnarray}
&&Z=  e^{A(\ell)} \int\!\!\!\int_< {\cal D}\psi^*{\cal D}\psi
e^{S[\psi^*,\psi]_<}\int\!\!\!\int{\cal
D}\bar{\psi}^*{\cal D}\bar{\psi} \,e^{S_0[\bar{\psi}^*,\bar{\psi}]} \,
\exp\Big\{\sum_{i=1}^4S_{I,i}[\psi^*,\psi,\bar{\psi}^*,\bar{\psi}]\Big\}\cr
&& \ \ \ =   e^{A(\ell)} \int\!\!\!\int_<
{\cal D}\psi^*{\cal D}\psi \exp\Big\{S[\psi^*,\psi]_<  +
\sum_{n=1}^\infty 1/n! \, \langle(
\,\sum_i
S_{I,i}[\psi^*,\psi,\bar{\psi}^*,\bar{\psi}]\,)^n\rangle_{\bar{0},c}\Big\}
\label{trace}
\end{eqnarray}
where
\begin{eqnarray}
\langle .... \rangle_{\bar{0},c} = Z^{-1}_{\bar{0}} \int\!\!\!\int{\cal
D}\bar{\psi}^*{\cal D}\bar{\psi} \, (.....) \,e^{S_0[\bar{\psi}^*,\bar{\psi}] }
\end{eqnarray}
is a free fermion average corresponding to a connected diagram evaluated in
the  outer momentum shell
and
\begin{equation}
Z_{\bar{0}}=\int\!\!\!\int{\cal D}\bar{\psi}^*{\cal D}\bar{\psi} \,
e^{S_0[\bar{\psi}^*,\bar{\psi}] },
\end{equation}
is the outer shell contribution to the free partition function.

\subsection{One-loop results}
The KW method may be used to obtain low order results presented in previous review
papers,\cite{Bourbon91,Bourbon95} and no particular technical difficulties arise. The following two
subsections present   material which has already been developed using the    parquet
summation\cite{Bychkov66,Dzyaloshinskii72} and the first order multiplicative renormalization 
group.\cite{Menyhard73,Kimura75,Solyom79}  Nevertheless, it is useful to supply the basic    properties of
the logarithmic theory of the   1D fermion gas model using this method.

\label{oneloop}
\subsubsection{Incommensurate band filling }
\begin{figure}[htb] \centerline{\includegraphics[width=10cm]{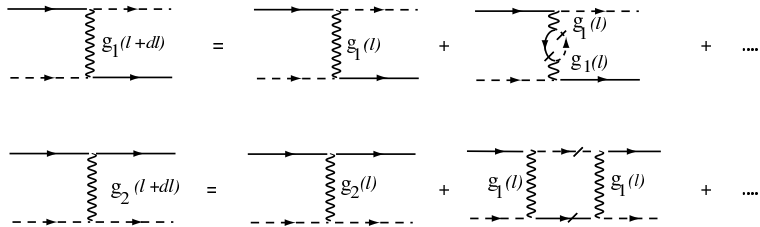}}
\caption{Recursion relations for backward  ($g_1$) and forward  ($g_2$)
scattering at the
one-loop level. The slashed lines refer  to a particle or a hole in the outer shell. 
}\label{recursion}\end{figure}

For   band filling that is incommensurate, there is no possibility of Umklapp
scattering so that one can drop $g_3$ in
the action. The renormalization of backward and forward scattering
amplitudes at the one-loop level  are
obtained from the  
$n=2$ outer-shell averages $\dm \langle (S_{I,2})^2\rangle$, where
\begin{eqnarray}
S_{I,2}[\psi ^{\ast },\psi,\bar{\psi }^{\ast },\bar{\psi } ]\  &=  &
S_{I,2}^C + S_{I,2}^P + S_{I,2}^L\cr 
&  \Longleftrightarrow &\
(\bar{\psi }_{+}^{\ast }\bar{\psi }_{-}^{\ast }\psi_{-}\psi
_{+} + {\rm c.c} ) + \, (\bar{\psi }_{+}^{\ast
}\psi^*_{-}\bar{\psi}_{-}\psi_{+} + {\rm c.c} )  \cr
&\ + &s  \ (\bar{\psi }_{+}^{\ast }{\psi }_{-}^{\ast }\psi_{-}\bar{\psi}
_{+} + {\psi }_{+}^{\ast }\bar{\psi }_{-}^{\ast }\bar{\psi}_{-}{\psi}
_{+}), 
\label{decomposition}
\end{eqnarray}
the three terms of which correspond to  putting two particles or two holes  in the outer
momentum shell (the Cooper channel),  one particle and
one hole on opposite branches   (the  Peierls 
channel), and finally putting a particle and a hole on the same  branch (the Landau channel). The latter
does not lead to logarithmic contributions at the one-loop level and can be ignored for the moment. The flow
of the scattering amplitudes   becomes $g_{i=1,2}(\ell + d\ell) = z_{i=1,2}(d\ell)g_{i=1,2}(\ell)$ where the
renormalization factors at the one-loop level are
\begin{eqnarray}
z_{1}(d\ell) &= & 1 - 2g_1(\ell)\,I_P(d\ell) +
2g_2(\ell) \,[I_P(d\ell)+
I_C(d\ell)]\cr 
z_2( d\ell) &= & 1 + g_2^{-1}(\ell)g_1^2(\ell)\,I_C(d\ell)
+ g_2(\ell)\, [I_P(d\ell)+
I_C(d\ell)].
\label{flowinc}
\end{eqnarray}
The outer shell integration  of the
Peierls channel at $2k_F$ and zero external frequency enables the particle and hole to be simultaneously
in the outer-shell. The result is 
\begin{eqnarray}
I_P(d\ell)&= & -\pi v_F{T\over L} \sum_{\{\kktil\}'}
G^0_+(k,\omega_n)G_-^0(k-2k_F,\omega_n)\cr
   &= & \left\{\int_{-\dm E_0(\ell)}^{-\dm E_0(\ell +d\ell)} \!\!\!+
\int_{\dm E_0(\ell +d\ell)}^{\dm
E_0(\ell)}\right\} {\tanh(\oq\beta E_0(\ell))\over E_0(\ell)}\, dE_0(\ell)\cr
 &  \simeq & \dm d\ell.
\label{pshell}
\end{eqnarray}
Similarly in the  Cooper channel at zero pair momentum and external frequency, we have
\begin{eqnarray}
I_C(d\ell)&= & \pi v_F{T\over L} \sum_{\{\kktil\}'}
G^0_+(k,\omega_n)G_-^0(-k,-\omega_n)\cr
   &= & -I_P(d\ell).
\end{eqnarray}
 Both contributions are logarithmic and their opposite signs  lead  to important
cancellations (interference). The 
diagrams  which remain  after the cancellation  are shown in Fig.~\ref{recursion}.

\begin{figure} \centerline{\includegraphics[width=7cm]{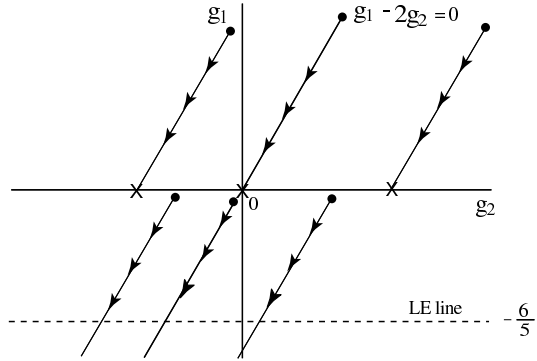}}
\caption{  Flow of coupling constants  in the $g_1g_2$ plane at the one-loop level. In the repulsive region
$g_1\ge 0$, the system has a line of fixed points on the Tomonaga-Luttinger line $g_1^*=0$.  In the
attractive sector
$g_1<0$, the  flows  is to  strong coupling. The dashed line corresponds
to the Luther-Emery line at
$g_1 =-6/5$.}\label{flow}\end{figure}

With an obvious
rearrangement, the flow equations may be written 
\begin{eqnarray}
&&{dg_1\over d\ell} =   -g_1^2 \cr
&&{d\over d\ell}(2g_2-g_1) =   0.
\end{eqnarray}
The solution for $g_1(\ell)$ is found at once to be 
\begin{equation}
g_1(\ell) = {g_1 \over 1+ g_1\ell},
\label{g1spin}
\end{equation}
showing that backward scattering is  marginally irrelevant (relevant) if
the bare coupling $g_1$ is
repulsive (attractive). As for the combination
$2g_2-g_1$, it remains marginal, an invariance that reflects the
conservation of particles on each branch.\cite{Fowler76,Rezayi81}
Another  feature of these equations is the fact that the flow of $g_1$  is
entirely uncoupled from
the one of $2g_2-g_1$.  This feature, which follows from  the
interference between
the Cooper and the Peierls channels,  gives rise to an
important property of   1D interacting fermion systems, namely the
separation of long wavelength spin and
charge degrees of freedom. This is rendered manifest
by rewriting  the interacting part of the action as follows\cite{Dzyaloshinskii72}
\begin{equation}
S_I[\psi^*,\psi]_\ell\ =  -\pi v_F(2g_2-g_1)\sum_{p,\qtil}
\rho_p(\qtil)\rho_{-p}(-\qtil) + \, \pi v_F\, g_1(\ell)
\sum_{p,\qtil}
   {\bf S}_p(\qtil)\cdot {\bf S}_{-p}(-\qtil)
\label{sc}
\end{equation}
where the long-wave length particle-density and spin-density fields of  branch $p$ are defined by
\begin{eqnarray}
\rho_p(\qtil)&=  &\dm\sqrt{T\over L}\sum_{\alpha,\{\kktil\}_<}
\psi^*_{p,\alpha}(\kktil +\qtil)
\psi_{p,\alpha}(\kktil)\cr
\ \ \ \ {\bf S}_p (\qtil) &=  &\dm\sqrt{T\over L}
\sum_{\alpha,\{\kktil\}_<} \psi^*_{p,\alpha}(\kktil +\qtil)
\, \vec{\sigma}^{\alpha\beta}\, \psi_{p,\beta}(\kktil).
\label{uniform}
\end{eqnarray}
The spin-charge separation is preserved at higher order and is a key
property of a Luttinger liquid in one
dimension.\cite{Haldane81} In the repulsive sector $g_1\ge0$, the
spin coupling goes to zero, while the
   forward scattering $g_2^*(\ell\to\infty) \to g_2-g_1/2$ flows to a
non universal value on the Tomanaga-Luttinger
line of fixed points $g_1^*=0$ (Figure~\ref{flow}).

   For an attractive interaction 
$g_1<0$, an instability occurs in the the spin sector at $\ell_\sigma$ defined by 
$g_1\ell_\sigma=-1$. Identifying $\ell$ with  $\ell_T=\ln E_F/T$,
the  temperature
 dependence of the Peierls  or Cooper loop, the characteristic temperature scale  for strong coupling is 
\begin{equation}
T_\sigma = E_Fe^{-1/\mid g_1\mid}\equiv \dm \Delta_\sigma,
\end{equation}
where $\Delta_\sigma$ is defined as the spin gap. According to (\ref{sc}),  backward scattering
can be equated with an exchange interaction
between  right- and left-going spin densities which is
antiferromagnetic in character when
$g_1$ is attractive. Therefore strong attractive coupling signals the
presence of a  spin gap
$\Delta_\sigma$. Although the divergence in (\ref{g1spin})  is an artefact of the
one-loop approximation, the flow of
$g_1(\ell)$ to strong coupling  is nevertheless preserved at
higher order. This is shown by the fact that the system
inevitably crosses the
so-called Luther-Emery line at
$g_1(\ell_{LE})=-6/5$ (Figure~\ref{flow}), where an exact solution using the technique of bosonization
confirms  the existence of a spin gap.\cite{Luther74b,Emery79}


\subsubsection{Umklapp scattering at half-filling}
At half-filling, there is  one fermion  per site so that $4k_F =2\pi/a$  coincides with a reciprocal lattice
vector $G$ enabling
Umklapp   scattering processes denoted by $g_3$ in Figure
\ref{Intg1g2}. The
presence of $g_3$ processes  reinforces the importance of the $2k_F$
density-wave channel compared to the
superconducting one. 
Flows 
will be modified accordingly. Thus at the one-loop level, the outer shell
corrections coming from $\dm \langle (S_{I,2})^2\rangle_{\bar{0},c}$ will
include additional contributions
\begin{eqnarray}
S_{I,2}    \Longleftrightarrow
S_{I,2}\big\vert_{g_3=0}&  +&(\,\bar{\psi }_{+}^{\ast }{\psi }_{+}^{\ast
}\bar{\psi}_{-}\psi_{-} + {\rm
perm.} + {\rm c.c}\,) \cr
 & + &(\,\bar{\psi }_{+}^{\ast }\bar{\psi }_{+}^{\ast
}{\psi}_{-}\psi_{-} +  \,{\psi }_{+}^{\ast }{\psi }_{+}^{\ast
}\bar{\psi}_{-}\bar\psi_{-}  + {\rm c.c}\,)
\end{eqnarray}
   involving outer shell fields in  the Peierls channel and   pairs of particles or holes on the same
branch. At the one-loop level only the former gives rise to logarithmic corrections, whereas the latter
will be involved in higher order corrections (\S~\ref{twoloopvert}). The former contributions, along with
(\ref{flowinc}), will lead to the recursion formulas $g_{i=1,2,3}(\ell + d\ell) =
z_{i=1,2,3}(d\ell)g_{i=1,2,3}(\ell)$, which can be written in form 
\begin{eqnarray}
   g_1(\ell + d\ell) & = &g_1(\ell) -\, 2g^{2}_1(\ell)I_P(d\ell), \cr
   \bigl(2g_2-g_1\bigr)(\ell + d\ell) & =& (2g_2-g_1\bigr)(\ell)  +\,
2g^{2}_3(\ell)I_P(d\ell), \cr
   g_3(\ell +d\ell) &  = &g_3(\ell) +
2\bigl(2g_2-g_1\bigr)(\ell) g_3(\ell)I_P(d\ell).
\end{eqnarray}
The flow of the coupling constants are then governed by
\begin{eqnarray}
{dg_1 \over d\ell}& = &-g^{2}_1, \cr
   {d \over d\ell}(2g_2 -g_1)& = &g^{2}_3, \cr
   {dg_3 \over d\ell}&=& g_3 (2g_2 -g_1),
\label{scalingu}
\end{eqnarray}
which corresponds to the diagrammatic parquet
summation of Dzyaloshinskii and Larkin. \cite{Dzyaloshinskii72} By combining the last two
equations, the flow is found to follow
the hyperbolas described by the renormalization invariant
   \hbox{$C= [(2g_2-g_1)(\ell)]^2 -g_3^2(\ell)$}.  Umklapp scattering
is then entirely uncoupled from $g_1$ and
will    only affect
   the charge
sector. When $g_1-2g_2 > \mid g_3\mid$, $g_3(\ell)$ scales to zero
and is  marginally irrelevant,
while
$2g_2^*-g_1^* $ is non universal. The excitation of charge degrees of freedom are then gapless in this
case.  The attractive Hubbard model in which $g_{1,2,3}=U<0$ falls in this
category.

\begin{figure} \centerline{\includegraphics[width=7cm]{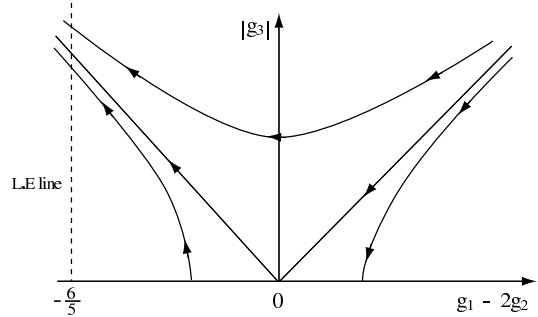}}
\caption{The flow of couplings in the charge sector. The condition
$g_1-2g_2 > \mid g_3\mid$ corresponds to weak
coupling where Umklapp scattering is a marginally irrelevant coupling. Otherwise,
 both $g_1-2g_2 $ and $g_3$ are
marginally relevant. The Luther-Emery line is
located at $g_1-2g_2=-6 /5$. }\label{flowcomm}\end{figure}

Otherwise, the flows of $g_3$ and
$2g_2-g_1$ evolve to strong coupling and both
  couplings are marginally relevant. There is a singularity at $\ell_\rho$ corresponding to the
temperature scale
\begin{equation}
T_\rho = E_F\, e^{-1/\sqrt{\mid C\mid}}\, \equiv \dm{\Delta_\rho}.
\end{equation}
 Following the example of the incommensurate case, the couplings
inevitably cross the Luther-Emery line at $g_1-2g_2=-6/5$, at which point
the problem can be  solved exactly and the
existence of a charge gap $\Delta_\rho$ is confirmed.\cite{Luther74b} 
The special case of the repulsive  Hubbard
limit at half-filling where
$g_{1,2,3}=U >0$ falls on the separatrix $2g_2-g_1= \mid g_3 \mid $
(Figure~\ref{flowcomm}). In this case the solution of (\ref{scalingu}) leads to a 
singularity at $T_\rho = E_F e^{-1/\mid
g_3\mid}$.  A  charge gap $\Delta_\rho$ is consistent with the
characteristics of a 1D Mott insulator found in
the Lieb-Wu exact solution. \cite{Lieb68,Dzyaloshinskii72,Larkin77}

\subsection{Two-loop results}
\label{Stwoloop}
   At the  one-loop level, corrections to the action parameters 
$\mu_S$ resulting from the partial trace are confined to the
scattering amplitudes
$g_{ i=1,2,3}(\ell)$ and aside from a renormalization of the chemical
potential, the  flow leaves the
single particle properties unchanged. At
the two-loop level,
 outer shell logarithmic contributions will affect  both the one-particle
self-energy and the four-point
vertices.  This yields an essential feature of the 1D electron gas model which is the absence of
quasi-particle states. 
\subsubsection{One-particle self-energy}
\label{selfsection}
\begin{figure}[htb] \centerline{\includegraphics[width=9cm]{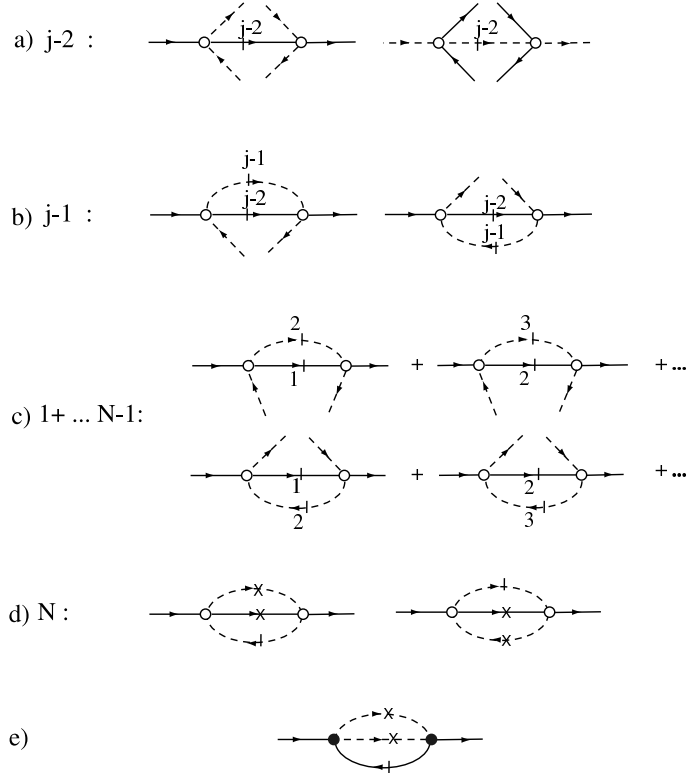}}
\caption{The cascade of contractions contributing to logarithmic one-particle
self-energy corrections in the local
approximation for the RG flow: (a) backward and forward scattering
contributions to the  
three-particle interactions; 
$\lambda^+$ (left) and
$\lambda^-$ (right) from $\dm \langle
S^2_{I,1} \rangle_{\bar{0},c}$  at the step $(j-2) $  where
$j\ge3$; (b) second contraction in the
outer shell at the step $j-1$ with two internal lines on successive energy shells  (only shown  for
$\lambda^+$)
; (c) the summation of diagrams given
in  (b) up to $j=N-1$   
  ; (d) outer shell one-particle self-energy correction  at the step $N\  (Nd\ell= e^{\ell +d\ell})$ for the
top inner shell state
$k$ of the 
$+k_F$ branch in  the local approximation. Here a crossed (slashed) line refers to fermions in the interval
$[0,\ell]$ ($[\ell,\ell+d\ell]$); (e) the result of a similar cascade of contractions for Umklapp
scattering.}\label{sixth}\end{figure}

  Let us
 consider  first the one-particle part. It is easy to see that the term 
$\dm\langle(S_{I,3})^2\rangle_{\bar{0},c}$ at $n=2$ has the form 
of  a one-particle self-energy correction.
  However, if we strictly follow the classical Wilson
scheme, the constraint of having
  {\it all}   internal
lines  in the outer shell leads to a vanishing 
 contribution.
    It becomes clear then that one-particle corrections can only be logarithmic at
the two-loop level if
the internal lines  refer to  different  momentum  shells.  In the classical Wilson scheme, this
possibility actually results from a cascade of
contractions  on  the three-particle interaction  (Figure \ref{sixth}). Such
interactions,
although absent in the
initial action, are generated along the RG flow and   enlarge
the parameter space.

In order to  see how these  interactions are generated from the partial
trace (\ref{trace}) and under what conditions  they give rise to logarithmic  corrections at the two-loop
level, it is useful to label the steps of the outer shell integration by the index $j$, for which
$E_0(\ell_j)= E_0e^{-(j-2)d\ell}$ is the scaled bandwidth at $j\ge 3$. Thus at each  step
$j-2$,  the contraction  $\dm \langle
S^2_{I,1} \rangle_{\bar{0},c}$  yields  a three-particle term   in the
action  (Figure
\ref{sixth}-a), which is denoted
$\delta S^{(j)}_\lambda$. Its explicit expression for the forward and
backward scattering is given in Eq.
(\ref{lambda}) of \S~\ref{six}. The analogous  expression $\delta S_{\lambda u}^{(j)}$ for Umklapp  with
$\lambda^\pm_u$ as the scattering
amplitudes is given in (\ref{sixthu}) of \S~\ref{sixu}. Owing to the presence of the internal
line, the rescaling of energy and fields shows that the amplitudes
$\lambda^\pm$  and  $\lambda_u^\pm$ are  marginal.\cite{Dupuis98} At the next step
 of partial integration $j-1$, two fermion fields of these  three-particle terms can in  turn  be
 contracted in  $\langle \delta
S^{(j)}_{\lambda,2}\rangle_{\bar{0},c}$, from which we  will retain an effective two-particle
interaction with two internal lines
tied to the adjacent shells
$j-2 $ and
$j-1$ (Figure~\ref{sixth}-b and (\ref{Sprime})). These contributions are non-logarithmic and
differ from those already
retained in the one-loop calculation in \S \ref{oneloop}  for which both
internal lines were put in the same outer shell. The repetition of these contractions is carried out by a
sum  on  $j$ up to $N-1$ (Figure~\ref{sixth}-c), which yields $\delta S'_\lambda\equiv \sum_j^{N-1}\langle
\delta S^{(j)}_{\lambda,2}\rangle_{\bar{0},c}$ ($
\delta S'_{\lambda,u}\equiv \sum_j^{N-1}\langle\delta S^{(j)}_{\lambda,u,2}\rangle_{\bar{0},c}$ for Umklapp)
as shown in (\ref{Sprime}) ((\ref{nonlocalu}) for
Umklapp). At the next step, $N$, of partial
integration, two more fields  are in turn   contracted which  lead  to a sum of one-particle self-energy
corrections for a $k$ state at the top of the inner shell.\cite{Note2} Since each term of the sum refers to
couplings at different steps $j$, the  correction is   non-local.
 
In the following, we will adopt a  {\it local} scheme of  approximation,  which consists in  neglecting the
slow (logarithmic) variation of the coupling constants in the intermediate integral over momentum transfer
$q$ in (\ref{local}) and evaluate them at
$\ell$  (Figure~\ref{sixth}-d). It is worth noting that the local approximation becomes exact for couplings
or combination of couplings that are invariant with
respect to $\ell$. This turns out to be the case for the Tomanaga-Luttinger model when only $g_2$ coupling 
is present ($g_1=g_3=0$), or for the combination
$2g_2-g_1$ when $g_3=0$. Neglecting small scaling deviations,
the end result   given
by (\ref{local}) ((\ref{localu}) for Umklapp) is
logarithmic and  contributes to a $1/z$ renormalization factor of $[G_p^0(k,\omega_n)]^{-1}$ for $k$ located
at the top of the inner shell. The  flow equation for  $z(\ell)$ is 
\begin{equation}
{d\over d\ell}\ln z = -{1\over 16} [(2g_2-g_1)^2 + 3 g_1^2 + 2g_3^2 ].
\label{flowself}
\end{equation}
The solution as a function of $\ell$ can be written as a product
$z=z_\sigma z_\rho $ of  renormalization
factors, in which
\begin{equation}
z_\sigma(\ell)  =  \exp\Big\{-{3\over 16}\int_0^\ell g_1^2 d\ell'\Big\}  
\end{equation}
for the spin part and 
\begin{equation}
z_\rho(\ell)  =   \exp\Big\{-{1\over 16}\int_0^\ell [(2g_2-g_1)^2 +
2g_3^2]d\ell'\Big\}
\end{equation}
for the charge. Prior  to  analyzing  the impact of these  equations on one-particle
properties, to work consistently at two-loop order, we must proceed to the
evaluation of the  two-loop contribution to the  coupling constant flow.

\subsubsection{Four-point vertices}
\label{twoloopvert}
\begin{figure}[htb] \centerline{\includegraphics[width=8cm]{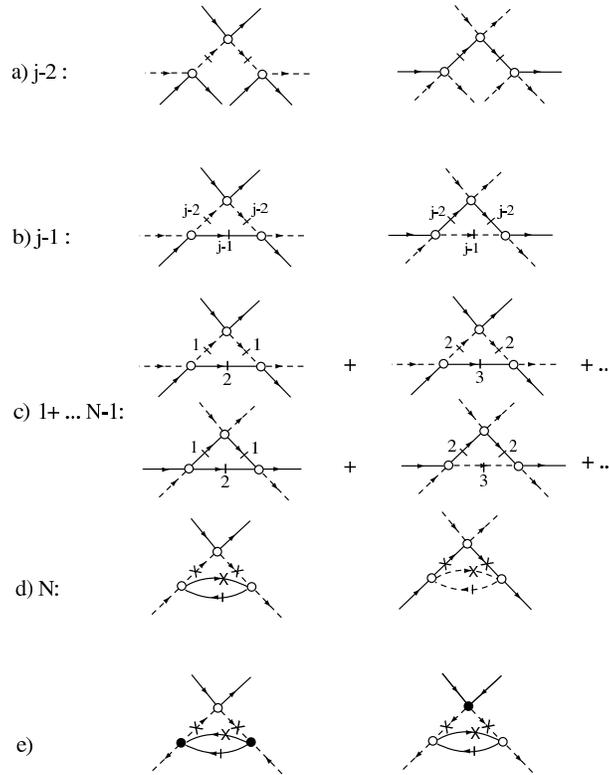}}
\caption{The cascade of contractions contributing to the four-point vertices in
the local
approximation of the RG flow: (a) backward and forward scattering
contributions to the generation of
four-particle interactions
 from ${1\over 2}\langle S_{I,2}^L S_{I,1}S_{I,1}\rangle_{\bar{0},c}$  at the step $j-2$ 
($j\ge3$) ; (b) second contraction in the
outer shell at the step $j-1$ ; (c) the summation of diagrams given
in  (b) up to $j=N-1$ as effective three-particle interactions; (d) two-loop outer shell correction to
four-point vertices in the local approximation at the step $Nd\ell\to \ell +d\ell$. Here a crossed  (dashed)
line refers to  fermion lines in the interval
$[0,\ell]$ ($[\ell,\ell+d\ell]$); (e) the result of a similar cascade of contractions
for Umklapp scattering.}\label{vertex}\end{figure}

To obtain the two-loop corrections to four-point vertices within  a
classical KW scheme, we come
up against  difficulties similar to those found for the one-particle self-energy. We can  indeed
verify that contractions
of the form $\langle S_{I,3}^2S^L_{I,2}\rangle_{\bar{0},c}$ at $n=3$, which
would lead to two-loop corrections
for the four-point vertices,  are actually vanishingly small if all internal lines  are in  the same
momentum shell. Here again non-zero
contributions  can only be found if the internal lines  refer to different
shells, which implies that the 
contractions must be made starting  from interactions that involve more than two  particles
(Figure~\ref{vertex}).\cite{Bourbon91}
Here we skip the technical details of the calculation and only sketch the main steps, which 
run parallel to those of   the 
one-particle self-energy.   The cascade of contractions starts from  the four-particle interaction  resulting
from 
${3\over 3!}\langle S_{I,2}^L S_{I,1}S_{I,1}\rangle_{\bar{0},c}$ at $n=3$, which involves the 
contribution $  S_{I,2}^L$ of the Landau channel.   The corresponding diagrams are shown in 
Figure~\ref{vertex}-a at the step
$j-2$. For the required contractions to be possible, the momentum transfer must be           zero.
At the next step, two more
fields are contracted and a line  is put in the adjacent shell $j-1$
(Figure~\ref{vertex}-b). As a function of
$j$, these corrections add up as shown in Figure~\ref{vertex}-c to lead to effective three-particle
interactions at the step $N-1$. At the last step,
$N$, of the cascade, one arrives at  a two-loop correction for the four-point vertices
(Figure~\ref{vertex}-d).  Following the example of one-particle self-energy, these corrections are
non-local in the coupling constants. In the  local approximation scheme, one finds an outer shell
contribution that is 
${\cal O}(g^3d\ell)$ (Figure~\ref{vertex}-d). The two-loop
contributions involving  Umklapp scattering are 
shown in Figure~\ref{vertex}-e. Finally, the outer shell contributions to the renormalization factors
$z_{1,2,3}$ (neglecting  corrections to logarithmic scaling at  $\ell$) are given
by
\begin{eqnarray}
z^{(2)}_1(d\ell) &= &1/4\Bigl(g^2_3(\ell) +
2 g_2(\ell)\bigl( g_2(\ell)- g_1(\ell)\bigr)\Bigr)d\ell, \cr
z^{(2)}_2(d\ell) & =& - g^{-1}_2(\ell)/4
\Bigl( g^3_1(\ell)- 2 g^2_1(\ell) g_2(\ell) +
2 g^2_2(\ell) g_1(\ell) \cr
 &  -&2 g^2_2(\ell)+
 g^2_3(\ell)\bigl( g_2(\ell)- g_1(\ell)\bigr)\Bigr)d\ell, \cr
z^{(2)}_3(d\ell) &=&1/4\bigl( g^2_1(\ell)
+2 g_1(\ell) g_2(\ell)-2 g^{2}_2(\ell)\bigr).
\end{eqnarray}

Now, combining these with the one-loop results of (\ref{scalingu}) and performing the
rescaling of the fields, one has 
$z^{-1/2} \psi^{(*)} \rightarrow \psi^{(*)}$
\begin{equation}
\mu_S=(G_p^0, z_1z^2g_1, z_2z^2g_2, z_3z^2g_3, \ldots),
\end{equation}
which renders explicit  the renormalization of  the couplings $g_i$. Neglecting scaling deviations at small
$\ell$, the corresponding scaling equations are 
\begin{eqnarray}
{dg_1 \over d\ell} & = &-g^{2}_1 -{1 \over 2}g^{3}_1, \cr
{d \over d\ell}(2g_2 -g_1) & =& g^{2}_3
[1-{1 \over 2}(2g_2 -g_1)],\cr
 {dg_3 \over d\ell} & = &g_3 (2g_2 -g_1)
[1-{1 \over 4}(2g_2 -g_1)]-{1 \over 4}g^3_3,
\label{twoloop}
\end{eqnarray}
 which reproduces the old results of the multiplicative renormalization group method. \cite{Solyom79,Kimura75}
We see that the spin coupling
$g_1$ is still  uncoupled to the charge couplings
$2g_2-g_1$ and
$g_3$. Furthermore, the one-loop singularities, previously found at $T_{\nu=\sigma,\rho}$,  are removed and 
replaced by  strong coupling fixed points for the couplings as $\ell
\rightarrow  \infty$. For the interesting cases already analyzed at the one loop level, we have   first
the attractive one where
$g_1<0$ and 
$g_1 - 2g_2 > \,\mid g_3 \mid $, for which $g_1^*\to -2$, $g_3^*\to 0$, while $g^*_1 - 2g^*_2$ is strongly
attractive but non universal;  second, in the repulsive case  when $g_1>0$  and $g_1 - 2g_2 <
\,\mid g_3
\mid
$,  for which    Umklapp 
is marginally relevant with  $g^{\ast}_3 \to 2, g^{\ast}_2 \rightarrow 1$ and
$g^{\ast}_1\rightarrow 0$. Since in both cases, the couplings cross the Luther-Emery line, the scale for
the  gap $\Delta_\nu$ remains of the order of the one-loop temperature scale $T_\nu$  in both sectors.

The integration of (\ref{flowself}) at large
$\ell
\gg
\ell_\nu$ then leads to   
  the power law decay
\begin{equation}
z(\ell) \sim \bigl(E_0(\ell)/\Delta_\nu\big)^{\theta^{\ast}}.
\end{equation}
 The exponent 
evaluated at the fixed point in the presence of a single gap can be separated into  charge ($\theta_\rho$)
and
spin ($\theta_\sigma$) components:
\begin{eqnarray}
\theta^{\ast} &&=\theta^{\ast}_\rho(2g^*_2-g_1^*,g^*_3)
+ \theta^*_\sigma(g^*_1) \cr
              &&\simeq 3/4.
\end{eqnarray}
 
The case when there is no Umklapp and $g_1>0$ corresponds to the Tomonaga-Luttinger model at large $\ell$. 
The couplings remain weak and one finds the power law decay 
\begin{equation}
z(\ell)=D_\sigma(\ell)\Bigl( E_0(\ell)/E_0 \Bigr)^{\theta_\rho},
\end{equation}
with
\begin{equation}
\theta_\rho={1\over 16}(2g_2 -g_1)^2,
\end{equation}
and
\begin{equation}
D_\sigma(\ell)=
\exp\Big(-{3 \over 16} \int^\ell_0 g_1^2(\ell')d\ell' \Big).
\end{equation}
Here  
the exponent $\theta_\rho$ is non-universal.\cite{Solyom79} A contribution from the spin
degrees of freedom appears through the weak
transient $D_\sigma(\ell)$. The physics of this gapless case is that of the Luttinger liquid. 

The factor $z$ is the quasi-particle weight at Fermi level. In all cases, it vanishes at
$\ell\to\infty$ or $T\to0$ indicating the absence of a Fermi liquid in one-dimension. Actually the same
factor also coincides  with the single-particle density of states at the Fermi level
\begin{equation}
N^*(\ell) = N(0) z(\ell),
\label{density}
\end{equation}
which  vanishes in the same conditions.

The connection between the one-particle Green's function computed at $\ell=0$ and that computed  at
non-zero
$\ell$ can be written in the form
\begin{equation}
G_p(\kktil,\mu_S)= z(\ell) G_p(\kktil,\mu_S(\ell)).
\end{equation}
When $\ell \to \infty $, the number of degrees of freedom to be integrated out goes to zero,
  thus
$G_p(\kktil,\mu_S(\ell))\to z G^0_p(k\to k_F,\omega_n)$ and as a function of $T\to 0$, one finds 
\begin{equation}
G_p(k_F,\omega_{n=0},\mu_S) \approx C{T}^{-\bar{\gamma}},
\end{equation}
where, consistent  with the scaling ansatz (\ref{scalingG}), $\bar{\gamma}= 1-\theta$ and  $C= 1/(i\pi
E_F^\theta)$. If we identify the vanishing energy $\dm E_0(\ell)$ in $z(\ell)$ with  $v_F(pk-k_F)$ close to
$k_F$, we have at
$T=0$ 
\begin{equation}
G_p(k,\mu_S) \propto \ \mid pk-k_F\mid ^{-1+\theta},
\end{equation}
which agrees with the scaling expressions (\ref{scalingGk}-\ref{Gvsx}) showing the absence of a Fermi liquid.
\subsubsection{Remarks}  
  At this  point, casting a glance back at the   RG approach described above,  a few remarks  can be made in
connection with the  alternative formulation presented in Refs. \cite{Bourbon91,Bourbon95}    In the  
classical Wilsonian scheme presented here,  a 
 momentum  cutoff $k_0$ (or bandwidth $E_0$)
  is imposed on  the spectrum at the start and all contractions done at  a given step of the RG refer  to
fermion states of the outer shell tied to that particular step. In this way, we have seen that   the two-loop
level calculations require a cascade of outer shell contractions starting from many-particle interactions
that were not present in the bare action. The cascade   therefore  causes non-locality of the flow
  which depends on the values of coupling constants obtained at previous steps. Thus
obtaining logarithmic scaling at higher than one-loop  forces us to resort to a local approximation.  

 In the formulation of Refs. \cite{Bourbon91,Bourbon95}, the Kadanoff transformation at high order is
performed differently. In effect, a  cutoff  $k_0$ is not  imposed on the range of momentum for
the spectrum in the free part of the action  $S_0$ but it only bounds the
 $k$-summations of the interaction term $S_I$. The latter summations, of which there two in both of the
interaction terms  (as in Eqn.~(\ref{partition}) ),  are involved in the ultraviolet
regularization of all singular diagrams, while the remaining summation on the momentum transfer $q$ is
treated   independently.    
  In this way, the interaction term $
S_I[\psi^*,\psi,\bar{\psi}^*,\bar{\psi} ]$ with fields to be integrated out in the  Kadanoff transformation
is defined differently than in (\ref{Sdecompos}). It has no $S_{I,1}$ term with a single field in the outer
shell, while   $S_{I,3}$ and $S_{I,4}$ contain a momentum transfer summation corresponding to the high-energy
interval
$\mid v_Fq\mid >E_0(\ell)/2$.  The
generation of new many-particle terms {\it via}
$S_{I,1}$ is therefore absent, and the evaluation of $\dm \langle (S_{I,3})^2\rangle_{\bar{0},c}$
for the one-particle self-energy correction, directly yields  (aside from   some differences    in the range
of  momentum  transfer integration and in the  scaling deviations) the end result of the local
approximation  of
\S~\ref{selfsection} at  the  two-loop  level. A similar conclusion holds for
the four-point vertices when the calculation  of  $\dm\langle S_{I,3}^2S^L_{I,2}\rangle_{\bar{0},c}$ is
performed.  Despite its simplicity, however, this alternative formulation is less
transparent  as  to the origin of logarithmic scaling in the 1D fermion gas model  at high order. Moreover,
  the presence of a summation over  a finite interval of momentum transfer   for each Kadanoff
transformation removes from the   fermion spectrum  its natural cutoff $k_0$ and   stretches  
the standard rules of the Wilson procedure.

\subsection{Response Functions}
\label{response}
We now  calculate  the response of the system to the formation of  pair correlations in the
Peierls and Cooper channels. In the Peierls channel, density-wave
correlations with charge or spin buildup centered either on sites or bonds differ at half-filling due to
the presence of  Umklapp scattering.  Actually, because $4k_F$ is a reciprocal lattice vector, density
waves at $+2k_F$ and  $-2k_F$ turn out to be  equivalent. In a continuum model, one can thus define the
following symmetric and  antisymmetric combinations of pair fields
\begin{equation}
O_{\mu^\pm_P}(\qtil)={1 \over {2}} \bigl(O^{\ast}_{\mu_P}(\qtil)
\pm O_{\mu_P}(\qtil)\bigr),
\end{equation}
corresponding to site ($+$) and  bond ($-$) 
(standing) density-wave correlations. 

\begin{figure}[htb] \centerline{\includegraphics[width=10cm]{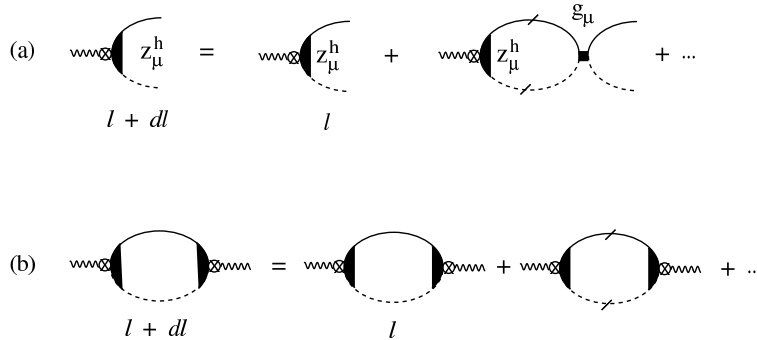}}
\caption{ Diagrams of the flow equations    of the pair vertex part $z_\mu^h$ (a) and susceptibility
$\chi_\mu$ (b) in the channel $\mu$. }\label{ChiRG}\end{figure}

We consider the linear coupling of these,  as well as  those previously introduced for the
Cooper channel in (\ref{singulet}-\ref{triplet}), to source fields $h_\mu$; the interacting part of the
action $S_I$ then  becomes
\begin{equation}
S_{I}\lbrack\psi^{\ast},\psi,h^*,h \rbrack =
S_{I}[\psi^{\ast},\psi] + S_{h}[\psi^{\ast},\psi],
\end{equation}
where
\begin{equation}
S_{h}[\psi^{\ast},\psi]=\sum_{\mu,\qtil}
\bigl(h_{\mu}(\qtil)O^*_{\mu}(\qtil) + {\rm c.c} \bigr).
\end{equation}
Following the substitution of the above $S_I$
in the action (\ref{action}), the KW transformation  will generate corrections  that are linear  in
$h^{(*)}_\mu$  and also terms with higher powers. In the
linear response theory, only linear and quadratic terms are kept so that the action at
$\ell$ takes the form

\begin{eqnarray}
S[\psi^{\ast},\psi, h^*,h]_{\ell}  &=&
S[\psi^{\ast},\psi]_{\ell}+\sum_{\mu,\qtil}
\bigl\{ z^h_{\mu}(\ell)
h^{\ast}_{\mu}(\qtil)O_{\mu}(\qtil) + {\rm c.c.} \cr
& -&\chi_{\mu}(\ell)h^{\ast}_{\mu}(\qtil)
h_{\mu}(\qtil)\bigr\},
\end{eqnarray}
where $z^h_{\mu}(\ell)$   is
the renormalization factor for  the pair vertex part  in the  channel $\mu$ (Fig.~\ref{ChiRG}-a). The term
which is quadratic in the source fields corresponds to the  response function (Fig.~\ref{ChiRG}-b)
\begin{equation}
\chi_{\mu}(\ell)=
-(\pi v_F)^{-1}\int^{\ell}_0 \bar{\chi}_{\mu}(\ell')d\ell',
\label{RGresponse}
\end{equation}
where 
\begin{equation}
\bar{\chi}_{\mu}(\ell) = \big(z_\mu^h(\ell)z(\ell)\big)^2
\label{barchi}
\end{equation}
is known as the auxiliary susceptibility of the channel $\mu$,\cite{Dzyaloshinskii72,Solyom79}  which is 
decomposed here into  a product of pair vertex and self-energy renormalizations. In the presence of source
fields, the parameter space of the action at the step $\ell$ then becomes
\begin{equation}
\mu_S(\ell)=(zG_p^0,z_1g_1,z_2g_2,z_3g_3,z_\mu^h,\chi _\mu,\ldots).
\end{equation}

 At the one-loop level $z=1$ and corrections to
$z^h_{\mu}$ come
from the outer shell averages $\langle S_{h}S_{I,2}\rangle_{\0til,c}$ for $n=2$.
Their evaluation is analogous to the one given in \S\ref{oneloop} and leads to the
  flow equation (Fig.~\ref{ChiRG})
\begin{eqnarray}
{d\over d\ell}\ln z^h_{\mu} &&=
{1\over 2}g_{\mu}(\ell),
\label{pairvertex}
\end{eqnarray}
where  $g_{\mu^\pm_P=0}=g_2 \mp g_3-2g_1$ and
$g_{\mu^{\pm}_P \not=0}(\ell)=g_2\pm g_3$ correspond to the   combinations  for $2k_F$ CDW and
SDW  site and bond correlations respectively, whereas
$g_{\mu_C=0}=-g_1-g_2$ and $g_{\mu_C\ne0}= g_1-g_2$ refer to couplings for SS and TS pair correlations. At
the two-loop level, the one-particle self-energy corrections contribute and from  (\ref{flowself}),
(\ref{barchi}) and (\ref{pairvertex}),  a flow equation  for the auxiliary susceptibility can be
written:
  \begin{equation}
{d\over d\ell}\ln\bar {\chi}_{\mu} =g_{\mu} - {1 \over
8}\,[\,(2g_2-g_1)^2 + 3 g_1^2 + 2g_3^2 \,],
\label{Aresponse}
\end{equation}
 which  coincides with the well known results of the multiplicative renormalization
group.\cite{Solyom79,Kimura75} In   conditions that lead to strong coupling, the use of the two-loop fixed
point values found in (\ref{twoloop})  
 yields the asymptotic power law form in temperature
\begin{equation}
\chi_{\mu}(T) \sim
\Bigl({T \over \Delta_\nu }\Bigr)^{-\gamma^{\ast}_{\mu}},\ \ \ \ \ \ \ {\rm for\ }\nu=\sigma,\rho.
\end{equation}
In the repulsive sector  $g_1>0$  and at half-filling, there is a charge gap ($\Delta_\rho\ne 0$). The
bond CDW, usually called   
 bond-order-wave (BOW), and site SDW correlations show a power law singularity with exponent
$\gamma^{\ast}_{BOW}=\gamma^{\ast}_{SDW}=3/2$,  as in the old calculations of the
multiplicative renormalization group.\cite{Kimura75} In the attractive sector, when $g_1<0$, and Umklapp
scattering is irrelevant, the gap is in the spins ($\Delta_{\sigma}\ne 0$ ) and both SS and incommensurate
CDW are  singular with the fixed point exponent $\gamma^*_{SS,CDW}=3/2$. (Here the SS response has a larger
amplitude.)\cite{Solyom79}
 However, these critical indices obtained from a two-loop RG expansion can only be considered as
qualitative estimates resulting from  fixed points, which are themselves artefacts
of the two-loop RG. Three-loop calculations have been achieved in the framework of multiplicative
renormalization group for
$g_3=0$, and
$\gamma_{SS}$ and
$\gamma_{CDW}$  turn out to be close to unity. The bosonization
technique on the Luther Emery line,\cite{Luther74b,Emery79,Voit95} allows an exact determination of these
indices at
$\ell
\gg \ell_{\nu}$ which are all equal to unity.

\begin{figure}[htb] \centerline{\includegraphics[width=6cm]{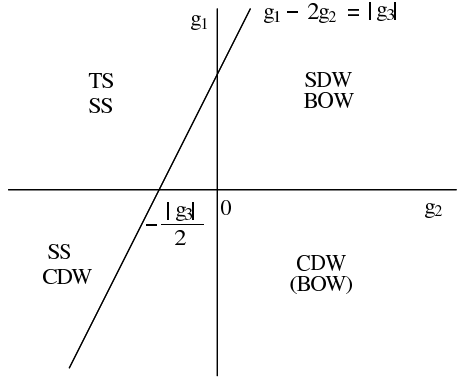}}
\caption{Phase diagram of the 1D interacting Fermi gas in the presence of Umklapp scattering.
}\label{1DPhases}\end{figure}

In the gapless regime for both spin and charge when $g_1>0$ and $g_1-2g_2>\mid g_3\mid$, the $\mu =$TS and
$\mu =$SS responses are singular  
\begin{equation}
{\chi}^{}_{\mu}(T)\ \sim
\Bigl({ T \over E_F }\Bigr)^{-\gamma^*_{\mu}}.
\end{equation}
Here the exponents $\gamma^*_{TS}=\gamma^*_{SS}=-g_2^*- g_2^{*2}/2$ are equal and take non universal values
on the Tomanaga-Luttinger line where only the attractive $g^*_2$ coupling remains at $\ell \to \infty$ 
(here the amplitude of the TS response is larger). Finally, there is a sector of coupling constants, that
is when
$g_1<0$ and $g_1-2g_2<\mid g_3\mid$, where a gap develops in both the spin and charge degrees of freedom. In
this sector, BOW (site CDW) is singular for $g_3>0$ ($g_3<0$) with an
exponent larger  than 2,\cite{Kimura75} which  indicates once again that the KW perturbative RG becomes
inaccurate in the strong coupling sector.  
 The  instabilities of the 1D interacting Fermi systems are
summarized in Figure
\ref{1DPhases}.
 Before closing this section we would like to analyze the scaling properties of the two-particle response
function. To do so, it is useful to restore the wave vector dependence of $\chi_\mu$ by
evaluating the upper  bound of the loop integral in (\ref{RGresponse}) at 
\begin{equation}
\ell\to \ell(q,T)= \ln {1.13 E_F\over T} + \, \psi(1/2) -
\dm\Big[\psi\Big(\, 1/2 + i{v_Fq\over 4\pi
T}\,\Big) + {\rm c.c }\, \Big],
\label{externalq}
\end{equation}
which coincides with the complete analytical structure of the free Peierls and Cooper loop in
(\ref{chiPeierls}) and (\ref{chiCooper}) in the static limit, respectively.  At  low temperature when $\mid
v_Fq\mid\gg 4\pi T\to 0$, we can use  (\ref{asymptoq}) and obtain
\begin{eqnarray}
\chi_\mu(q) \propto q^{-\gamma_\mu} \hskip 1 truecm {\rm or} \hskip 1 truecm \chi_\mu(x) \propto
x^{\gamma_\mu-2} , 
\end{eqnarray}
which are consistent with the scaling relation $ 2-\eta_\mu= \gamma_\mu$  at $\nu=1$
(Eqn.~(\ref{scalinglaw})).

\section{Interchain coupling: one-particle hopping}

Let us now consider the effect of a small interchain hopping (Figure~\ref{hopping}) in the framework of the
KW renormalization group. The bare propagator of  the action at $\ell=0$ is then replaced by
$G_p^0(\kvec,\omega_n)$ given in (\ref{tperp}). The parameter space for the action at $\ell=0$ then
becomes 
$$
\mu_S=(G_p^0(k,\omega_n),t_\perp,g_1,g_2,g_3).
$$ 
Now applying the  rescaling transformations 
(\ref{rescaleE}-\ref{rescaleF}) in the absence of interaction,    
 the 
rescaling   of 1D energy leads to the transformation 
\begin{equation}
t'_{\perp}=st_{\perp},
\end{equation}
 which means that
the canonical dimension of $t_{\perp}$ is unity  and  it is therefore a
{\it relevant} perturbation.  An estimate of the temperature at which a dimensionality crossover in
the coherence of single-particle motion occurs can be obtained by replacing $s$   with the ratio 
$\xi/a$  and setting 
$t'_{\perp}\sim E_F$. We find $T_{x^1} \sim t_{\perp}$ in agreement with
the results of \S~\ref{sechopping}.  How this is 
 modified in the presence of interactions will depend on the anomalous dimension
$\theta$ which  the fermion field  acquires under 1D renormalization.  When the partial trace is
performed at   low values of $\ell$, one-particle self-energy corrections are essentially 1D in
character and are governed by (\ref{flowself}). At  step
$\ell$, the effective 2D propagator then takes the form
\begin{equation}
G^{0}_{p}({\bf k},\omega_n)=z(\ell)\lbrack  G^{0-1}_p(\kktil)+
2t_{\perp} z(\ell)\cos k_{\perp} \rbrack^{-1}.
\label{FLpropagator}
\end{equation}
Therefore  the power law decay of the quasi-particle weight and the density of states (cf.
Eqn.~(\ref{density})) along the chains  leads to a renormalized hopping amplitude
$zt_{\perp}$ which decreases as a function of $\ell$.  The  condition for the single particle
crossover temperature $T_{x^1}$ is then  renormalized, becoming $z(T)t_\perp\xi/a\sim E_F$, which implies 
\begin{eqnarray}
T_{x^1}  & \sim & t_{\perp }
\bigl({t_{\perp } \over E_F} \bigr)^{{1-\theta \over \theta}} \cr
&  \propto& t_{\perp }^{1/\phi_{x^1}},
\end{eqnarray}
where $\phi_{x^1} = 1-\theta$ is the single
particle crossover exponent. \cite{Bourbon84,Bourbon91} 
In this simple picture, well below $T_{x^1}$, the system behaves  as a Fermi liquid. The  effective bare
propagator (\ref{FLpropagator}) for $E_p({\bf k}) < \dm E_0(\ell_{x^1})$, where $\ell_{x^1}=\ln E_F/T_{x^1}$,
has the Fermi liquid form (\ref{scalingG}) with a weight that is consistent with the extended scaling
expression (\ref{coefscale}) with $\bar{\gamma}=1-\theta$,
$\dot{\bar{\gamma}}=1 $ and the non universal constant $z_\infty = E_F^{\theta/(\theta-1)}$.

\begin{figure} \centerline{\includegraphics[width=7cm]{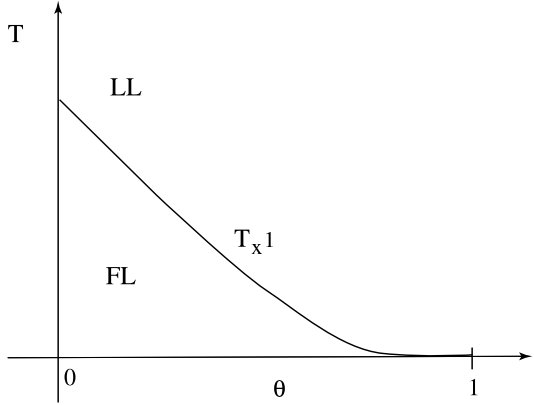}}
\caption{Crossover temperature variation as a function of the one-particle self-energy exponent $\theta$. 
}\label{Txprofile}\end{figure}

As
   long as $\theta < 1$, $t_{\perp}$ remains a relevant variable and
$T_{x^1}$ is finite. For $\theta = 1$ and $\theta >1 $, however, it becomes
marginal and irrelevant respectively.  In these latter cases,  transverse band
motion has no chance to develop and fermions are confined along the
chains at all temperatures.  In the KW framework, conditions for 1D single-particle confinement are
  likely to be 
realized in the presence  of  a gap  since in this case
  we have seen that
$\theta \sim
1$. \cite{Brech90,Voit98,Carlson00}   Single-particle confinement is also present  
in the absence of any gaps, as in the
Tomanaga-Luttinger model, where $g_1=g_3=0$ and $\theta \sim 1$ when $\mid
g_2 \mid$
is close to unity \cite{Emery79,Luther74b} and  other interacting fermion models
\cite{Capponi98,Vishwanath01}. The dependence of the single-particle crossover temperature on
the interaction strength,  parametrized by the exponent
$\theta$, is given in Figure~\ref{Txprofile}.

A point worth stressing here is   that the
 expression (\ref{FLpropagator})  for the effective propagator 
is equivalent to an RPA approach for the interchain motion and is therefore not exact.
\cite{Boies95,Arrigoni98,Caron88,Biermann01} It  becomes exact when interchain single-particle hopping has an
infinite range.\cite{Boies95} Another point worth stressing is the fact that  the temperature interval over
which the single-particle dimensionality crossover takes place  is not small. In effect, as already
noticed in
\S~\ref{Scaling1D},  the critical 1D  temperature domain stretches from $E_F$  to $T$  making the crossover
as a function of temperature  gradual.   

\subsection{Interchain pair hopping and long-range order}
\label{Spairhopping}

Although transverse coherent  band motion is absent above $T_{x^1}$,
   a finite probability for fermions to make a jump on neighboring
chains still survives  within  the thermal coherence time $\tau \sim 1/ T$.
When this  ``virtual'' motion is combined with electron-electron
interactions along the stacks, it generates an effective interchain hopping for
{\it pairs}  of particles that are created in all 1D channels of correlations
(Peierls, Cooper, and Landau).\cite{Bourbon86,Bourbon88,Bourbon91}  When coupled with
singular pair correlations along the chains, as  may occur in the Cooper or Peierls channels, effective
transverse two-particle interaction may arise and lead to the occurrence of long-range order above 
$T_{x^1}$, even when  fermions are still confined along the chains  either by thermal fluctuations or by
the presence of a gap. In other words, interchain pair processes generate  a different kind of
dimensionality crossover, which in turn introduces  a new  fixed point connected to long-range order. It
should be noticed, however, that  long-range order  is not possible in two dimensions
if the order parameter has more than one component.\cite{Mermin66}  In such a case, a finite coupling  in a
third direction is required.  
 
\begin{figure} \centerline{\includegraphics[width=9cm]{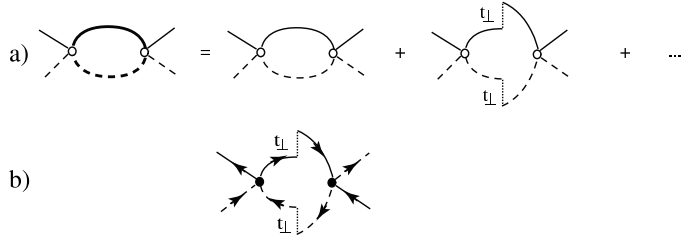}}
\caption{ a) Generation of interchain pair hopping from one-loop backward and forward vertex 
corrections in the Cooper and Peierls channels at each outer shell integration. Here the thick (thin)
continuous and dashed lines refer to
$\pm k_F$  2D (1D) propagator. b) The Umklapp contribution to interchain  pair hopping in the Peierls
channel.  }\label{generator}\end{figure}

To see how these new possibilities emerge  from the
renormalization group method described above,  the  quasi-1D
propagator (\ref{FLpropagator}) is used   in the evaluation of the one-loop outer shell corrections 
 $\dm\langle S^{2}_{I,2}\rangle_{\0til,c} $ and  the perturbative influence
of $t_{\perp}$ in both Peierls and Cooper channels
 is shown in Figure~\ref{generator}. The first term  is independent of
$t_{\perp}$ and reproduces  the 1D one-loop  corrections. 
The second diagram expresses the generation at each outer shell
integration of an elementary contribution to interchain  hopping
 of a pair of particles in the Peierls or Cooper channel.  This
outer shell generating term is given by
\begin{equation}
S^{0}_{\perp}[\psi^{\ast},\psi]= - {d\ell \over 2} z^{-2}(\ell)\pi v_F
\sum_{\mu} \sum_{\QQtil} f^{}_{\mu}(\Qvec,\ell)
O^{ \ast}_{\mu}(\QQtil)O_{\mu}(\QQtil),
\label{pair}
\end{equation}
where $\Qvec=(q_0,q_\perp), \QQtil=(q,q_\perp,\omega_m)$ and 
\begin{equation} 
f_{\mu}(\Qvec,\ell)= \pm2
\Bigl[ {g_{\mu}(\ell) z(\ell)t_{\perp } \over
E_0(\ell)} \Bigr]^{2}\cos q_{\perp }.
\end{equation}
The upper (lower) sign refers to the Peierls  (Cooper) channel.

\begin{figure} \centerline{\includegraphics[width=7cm]{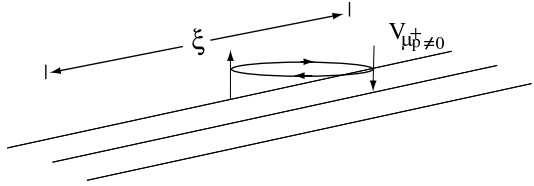}}
\caption{Antiferromagnetic interchain exchange $V_{\mu_P^+}$ occurring between two spins within the coherence
length
$\xi$. In the presence of a charge gap, the exchange takes place in the interval $\xi\to\xi_\rho\sim
v_F/\Delta_\rho$. }\label{Exchange}\end{figure}

This  term was not present in the bare action and must be added to
$\mu_S(\ell)$.  Furthermore, it will be coupled to other
terms of $S$ giving rise to additional corrections and hence to an
effective interchain two-particle action of the
form
\begin{equation}
   S_{\perp}[\psi^{\ast},\psi]= -{1 \over 2} z^{-2}(\ell)
\sum_{\mu} \sum_{\QQtil} V_{\mu}(\Qvec,\ell)
O^{ \ast}_{\mu}(\QQtil)O_{\mu}(\QQtil),
\label{Vmu}
\end{equation}
where $V_{\mu}(\Qvec,\ell)$ is the  pair
tunneling amplitude. As a function of $\ell$, the parameter space of the action then becomes
$$
\mu_S(\ell)=(zG_p^0,zt_\perp,z_1g_1,z_2g_2,z_3g_3,V_\mu,\ldots)
$$
Among the most interesting pair tunneling processes in the Peierls channel,
$V_{\mu^+_P\not=0}$ corresponds to an interchain kinetic exchange and promote site SDW or AF ordering
(Figure~\ref{Exchange}), while 
$V_{\mu^-_P=0}$ is an interchain BOW coupling which is involved in   spin-Peierls
ordering. \cite{Bourbon95} Both favor the transverse
propagation of order at
$\Qvec_{0}=(2k_F,\pi)$, which actually coincides with
the best nesting vector
of the whole {\it warped}  Fermi surface. In the Cooper channel, $V_{\mu_C}$  describes
the interchain Josephson coupling for SS and TS Cooper pair hopping. According to 
Figure~\ref{1DPhases}, these terms can only couple to singular
intrachain correlations when Umklapp scattering is irrelevant.  Long-range order  in
the Cooper channel is uniform and occurs at the wave vector
$\Qvec_0=(0,0)$.

\begin{figure} \centerline{\includegraphics[width=14cm]{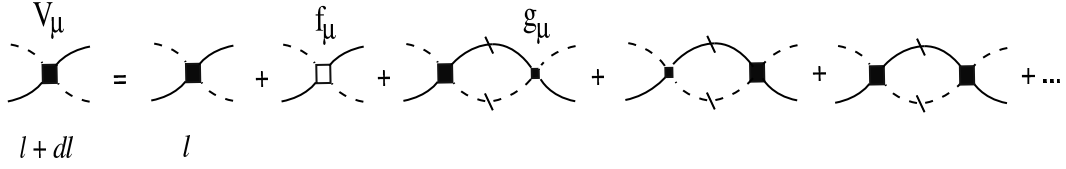}}
\caption{Recursion relation for the interchain pair hopping term in the transverse RPA treatment
(Eqn.(\ref{flowrpa})). }\label{recurenceP}\end{figure}

Long-range order will occur  at the critical point determined by $\ell_c =
\ln(E_F/T_{c,\mu})$ where $V_{\mu}(\Qvec_0,\ell_c)\rightarrow
\infty$.  A reasonable way to obtain an approximate   $T_{c,\mu}$ is a
transverse RPA approach (Fig~\ref{recurenceP}), which consists of treating interchain coupling in the
mean-field approximation while the intrachain correlations are taken into account rigorously using
(\ref{Aresponse}) or the exact asymptotic form of the pair susceptibility when it is known.  Thus substituting
$S_I
\rightarrow S_I + S_{\perp}$ in (\ref{trace}), one gets the one-loop outer-shell contribution ${1 \over
2}\langle S^{2}_{I,2} + 2S_{I,2}S_{\perp,2} + S^{2}_{\perp,2} \rangle_{\0til}$.  The first term leads to
the generating term (\ref{pair}) to second order in
$zt_\perp$, whereas  the second term introduces pair vertex 
corrections to the composite field $O_{\mu}^{(*)}$ which can then be expressed in
terms of $\bar {\chi}_{\mu}$ given in (\ref{Aresponse}).  Finally, the third term is the RPA
contribution that enables the free propagation of  pairs in the
transverse direction and hence the possibility of long-range order at finite
temperature (Fig.~\ref{recurenceP}).  In all cases other than the generating term, the outer-shell
integration in the  RPA approach is made by neglecting $t_\perp$, with the result
\begin{equation}
{d\over d\ell}V_{\mu}(\Qvec,\ell) = f_{\mu}
+ V_{\mu}{d \over d\ell}\ln\bar{\chi}_{\mu}
- {1 \over 2} (V_{\mu})^2 
\label{flowrpa}
\end{equation}
for the flow equation of the interchain pair hopping. The solution of
this inhomogeneous equation at $\ell$ has the  simple pole structure
\begin{equation}
V_{\mu}(\Qvec,\ell)={\bar{V}_{\mu}(\Qvec,\ell) \over
1-{1 \over 2} \bigl( \bar {\chi}_{\mu}(\ell)\bigr)^{-1}
\bar {V}_{\mu}(\Qvec,\ell)
\chi_{\mu}(\ell)},
\label{VRPA}
\end{equation}
where
\begin{eqnarray}
\bar{V}_{\mu}(\Qvec,\ell)=\bar {\chi}_{\mu}(\ell)
\int^{\ell}_{0} &   d\ell' &f_{\mu}(\Qvec,\ell')
\bigl( \bar {\chi}_{\mu}(\ell')\bigr)^{-1} \cr
& \times& \bigl[ 1-{1 \over 2} \bigl( \bar {\chi}_{\mu}(\ell')\bigr)^{-1}
\bar {V}_{\mu}(\Qvec,\ell')
\chi_{\mu}(\ell')\bigr]^2.
\label{Vbar}
\end{eqnarray}

\subsubsection{Critical temperature in the random phase approximation}
\label{criticalTc}

\begin{figure} \centerline{\includegraphics[width=7cm]{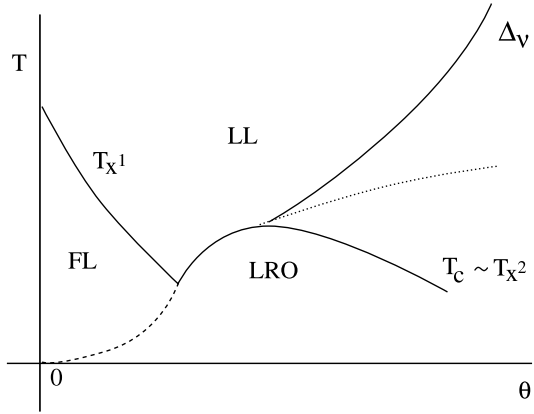}}
\caption{Schematic diagram of  the temperature ($T$)  and interaction  strength ($\theta$) dependence
of the single-particle  and pair dimensionality crossovers   from the Luttinger liquid
(LL) or Luther-Emery liquid ($T<\Delta_\nu$) towards a Fermi liquid (FL) or long-range ordering (LRO). The
dashed (dotted) line corresponds to the onset of LRO from a Fermi liquid (gapless LL in strong coupling).
}\label{Tcprofile}\end{figure}

Owing to the structure of $\bar{V}_{\mu}$, a general
analytic  solution
for $T_{c,\mu}$ is impossible. However, an approximate expression for the
critical
temperature can be readily obtained, by first neglecting transients in
the power laws of $\bar {\chi}_\mu\sim e^{\gamma_\mu\ell}$, $z\sim e^{-\theta \ell}$, and  the couplings
$g_\mu(\ell)$, and by setting to unity the term in  square  brackets in 
the integrand of (\ref{Vbar}).  Thus,  far from $T_{c,\mu}$ where the RPA
term can be
neglected, one can write
\begin{equation}
V_\mu(\Qvec_0,\ell)\approx 2{\langle g^2_{\mu}\rangle t^2_{\perp } 
\over E^2_0}\bigl((2-2\theta-\gamma_\mu)^{-1}
(e^{\ell(2-2\theta-\gamma_\mu)}-1\bigr),
\label{Veffective}
\end{equation}
 where $\langle g_{\mu}^2\rangle$  is an average
of the coupling constants over the interval $[0,\ell]$. 
 In the presence of a gap in the charge or spin sector, the
above scaling form at $\ell\gg\ell_{\nu=\rho,\sigma}=\ln(E_F/T_{\sigma,\rho})$ is governed by the strong
coupling condition
$2-2\theta^{*}-\gamma^{*}_{\mu}< 0$ , while for 
$\ell\ll\ell_{\nu=\rho,\sigma}$, the couplings are weak and $2-2\theta-\gamma_{\mu}> 0$.  In strong
coupling when 
$\ell\mid 2-2\theta^{*}-\gamma^{*}_{\mu}\mid  \gg 1$, the expression (\ref{Veffective})
actually coincides with $\bar{ V}_\mu$ so that its substitution in (\ref{VRPA}) leads
to a non-self-consistent pole at
\begin{equation}
T_{c,\mu} \approx T_{\nu} \bigl[{g_{\perp\mu} \over \delta_\mu
\gamma^{*}_{\mu} }\big(1+ {g_{\perp\mu}\over\delta_\mu
\gamma^{*}_{\mu}}\big)^{-1} \bigr]^{1/\gamma^{*}_{\mu}},
\end{equation}
where
\begin{equation}
g_{\perp\mu}= {\langle g_{\mu}^2\rangle
z^2(\ell_\nu)\over
2-2\theta-\gamma_{\mu}}
{t^{2}_{\perp } \over \Delta^{2}_{\nu}} 
\end{equation}
is an effective interchain pair hopping term
at the step
$\ell_\nu$. This form is
reminiscent of a perturbative expansion in terms of
the small parameter $zt_{\perp}/\Delta_{\nu}$ indicative of interchain
hopping of bound pairs.
\cite{Firsov85,Brazovskii85a,Suzumura85} The
positive constant
\begin{equation}
\delta_{\mu} = 1+  g_{\perp\mu}(\Delta_{\nu}/E_0)^{\gamma_{\mu}}
\chi_{\mu}(\Delta_\nu) <1 
\end{equation}
  corresponds to the integral of   $(\ref{Vbar})$ over the interval $[0,\ell_\nu]$ at high energy. 
These results  in the presence of a gap allow us to write 
\begin{eqnarray}
T_{c,\mu} \sim \  T_{x^2} \sim && \ g_{\perp\mu}^{1/\phi_{x^2}},
\end{eqnarray}
which is compatible with
(\ref{cross2}) from  the extended scaling ansatz.  $T_{x^2}$ is thus a temperature scale for the
dimensionality crossover for the correlation of {\it  pairs } of particles  induced by
$g_{\perp\mu}$, and $\phi_{x^2}=\gamma_\mu ^* $ is the two-particle exponent.\cite{Bourbon91} $T_{x^2}$ may
 also be equated with the onset of critical (Gaussian) fluctuations.
A point that
is worth noticing here concerns  the variation of the critical temperature with the gap $\Delta_\nu$. 
Actually, for most interesting situations, the flow of coupling constants will cross the
Luther-Emery line and  the use of the expression (111) with
$\gamma^*=1$  will lead to the dependence $T_{c,\mu}\approx z^2(\ell_\nu)t^2_\perp/\Delta_\nu$ in  which
$T_{c,\mu} $ increases as the amplitude of the gap decreases (Figure~\ref{Tcprofile}).  
 In the absence
of a  gap, the interchain pair hopping term is still present and it can lead to long-range
ordering. Sufficiently far from the critical point and for $\ell(2-2\theta-\gamma_{\mu}) \gg 1$, one
finds 
\begin{eqnarray}
V_{\mu}(\Qvec_0,\ell)&= & \bar{V}_{\mu}(\Qvec_0,\ell) \cr
&\approx  &  -2g_{\perp\mu}(2-2\theta-\gamma_\mu)^{-1}
(E_{0}(\ell)/E_0)^{2\theta-2}
\end{eqnarray}
where in weak coupling  $g_{\perp\mu}= \langle g_{\mu}^2
\rangle t^{2}_{\perp }/E^2_0$.  Substituting this expression for $\bar{V}_{\mu}$ in the RPA
expression (\ref{VRPA}), the critical point will occur at
\begin{equation}
T_{c,\mu} \sim E_F\bigl
( g_{\perp\mu}
/[\gamma_{\mu}(2-2\theta-\gamma_{\mu})] \bigr)^{{1 \over 2(1-\theta)}}.
\label{Tcweak}
\end{equation}
 Therefore, in weak coupling,  $T_{c,\mu} \sim T_{x^2} \sim
(g_{\perp\mu})^{1/\phi_{x^2}}$, and  one can define the two-particle crossover
exponent  $\phi_{x^2}=
2(1-\theta)$ which is twice $\phi_{x^1}$ (plus anomalous corrections coming from the $g_\mu$). The critical
temperature
$T_{c,\mu}\approx 
\langle g_{\mu}^2
\rangle^\dm \,zt_\perp$ will then decrease when interaction decreases. 
This result is meaningful as long as $T_{x^2} > T_{x^1}$, that is
when the  two-particle dimensionality crossover  that marks
the onset of Gaussian critical fluctuations still falls in a temperature region where
the transverse
single particle motion is incoherent. The interval of coupling constants where this occurs is relatively
narrow, however, since when there is no gap $t_\perp$ is a relevant parameter and the calculated $T_{x^1}$
increases  with decreasing interactions. When the strong and weak coupling regimes are looked at together, 
a maximum of $T_{c,\mu}$ is therefore expected at the boundary (Figure~\ref{Tcprofile}). 

Finally, it should be stressed that whenever the strong coupling condition $2-2\theta-\gamma <0$ is achieved
in the absence of any gaps as   in the case for the Tomanaga-Luttinger model, there is no maximum of
$T_{c,\mu}$. Instead, it is found to increase  monotonically with the interaction strength (see the dotted
line of Figure~\ref{Tcprofile}).\cite{Boies95,Brazovskii85a}

\subsubsection{Response functions }

The calculation of the response functions in the presence of interchain pair hopping can be done following
the procedure given in \S~\ref{response}.  At the RPA level for the pair hopping, one must add the
outer-shell contribution $\langle S_hS_{\perp,2}\rangle_{\bar{0},c}$ to the scaling equation for
$\bar{\chi}_\mu$ in  the 1D  regime (see Eqn.~(\ref{Aresponse})),  to get
\begin{equation}
{d\over d\ell}\ln\bar{\chi}_\mu(\Qvec,\ell) = {d\over d\ell}\ln\bar{\chi}_\mu(\ell)- V_\mu. 
\end{equation}
Using (\ref{VRPA}), the integration of this equation leads to
\begin{equation}
\bar{\chi}_\mu(\Qvec,\ell)\approx {\bar{\chi}_\mu(\ell)\over \big[\, 1-{1 \over 2}
\bigl(\bar{\chi}_{\mu}(\ell)\bigr)^{-1}
\bar{V}_{\mu}(\Qvec,\ell)\chi_\mu(\ell)\,\big]^2},
\end{equation}
which presents a  singularity that is quadratic in $(T-T_{c,\mu})^{-1}$   at
$\Qvec=\Qvec_0$. The exponent
$\dot{\gamma}$ (cf. Eqn.~(\ref{chitc})) for the auxiliary susceptibility is therefore twice the value
expected for 
$\chi_\mu$ in the transverse RPA approach. For the  cases of two-particle crossovers in \S~\ref{criticalTc},
this expression at
$\Qvec_0$ can also be written in an extended scaling form similar to (\ref{chiX})
\begin{equation}
\bar{\chi}_\mu(\Qvec_0,\ell)\approx\bar{\chi}_\mu(\Qvec_0,\ell)X ( B g_{\perp\mu }/T^{\phi_{x^2}})
\end{equation}
 where $X(y)=1/(1-y)^2$ is the RPA crossover scaling function for the auxiliary
susceptibility.
  
In the present transverse RPA scheme, the small longitudinal $q$ dependence can be  incorporated in
$\bar{\chi}_\mu$ through the substitution (\ref{externalq}), while the $q_\perp$ dependence will come from
 an expansion of $\bar{V}_\mu$ in (\ref{Vbar}) around the transverse ordering wavector $q_{\perp0}$ in the
channel considered. The {\bf Q} dependence of $\bar{\chi}_\mu$ will take   the form
\begin{equation}
\bar{\chi}_\mu(\Qvec,T) \propto  (\dot{t} + \dot{\xi}_0^2q^2 +\dot{\xi}_{\perp0}^2 q_\perp^2)^{-2}
\end{equation}
where  $\dot{\xi}_0\sim v_F/T_c$ is the longitudinal coherence length; the transverse coherence
length $\dot{\xi}_{\perp0}< 1$ is smaller than the interchain distance  indicating strong anisotropy. The
growth of the correlation length is governed by the usual Gaussian expression $\vec{\dot{\xi}}=
{\dot{\!\vec{\xi}}_0}\,\dot{t}^{-1/2} $, which according to (\ref{correlation}), leads to the correlation
exponent
$\dot{\nu}=1/2$. Similar  agreement with  scaling  is found for the response function, for which the
above RPA treatment leads to the familiar results $\dot{\gamma}=1 $ and
$\dot{\eta}=0$.
\subsection{Long-range order in the deconfined region}
\subsubsection{Critical temperatures in the ladder approximation}
\label{BelowTx1}
\begin{figure} \centerline{\includegraphics[width=10cm]{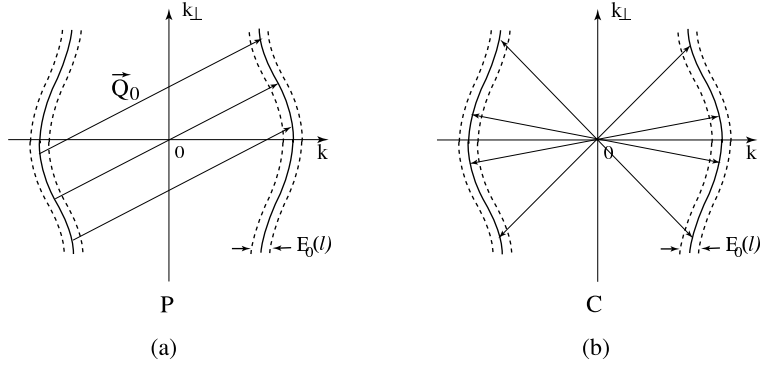}}
\caption{ (a) particle-hole
pairing in the Peierls channel at $\Qvec_0=(2k_F,\pi)$. (b) Particle-particle (or  hole-hole) pairing in the
Cooper channel. Constant energy surfaces at $\pm\dm E_0(\ell>\ell_{x^1})$ from the Fermi surface are shown as
dashed lines. }\label{Fsurface}\end{figure}

When the calculated value of $T_{x^2}$ becomes smaller than  the single-particle deconfinement  temperature 
$T_{x^1}$, the two-particle crossover is irrelevant and the expression (\ref{Tcweak}) for $T_{c,\mu}$ becomes
incorrect. In effect, thermal fluctuations have energies smaller than the characteristic energy
related to the warping  of the Fermi surface, which then becomes coherent. The
$t_\perp$ expansion of Fig.~\ref{generator}  stops being convergent and  the flow equation
(\ref{flowrpa}) must be modified accordingly.  Now
the effective low-energy  spectrum 
\begin{equation}
E_p(\kvec)= \epsilon_p(k) -2zt_\perp\cos k_\perp
\end{equation}
 is characterized by the electron-hole  symmetry (nesting)  $E_+(\kvec+\Qvec_0)=-E_-(\kvec)$
and   logarithmic corrections will also be  present below $E_0(\ell_{x^1})$ (Figure~\ref{Fsurface}-a).
Similarly in the Cooper channel, the inversion property of the spectrum $E_+(\kvec)=E_-(-\kvec)$ also leads
to a logarithmic singularity in the same energy range (Figure~\ref{Fsurface}-b). We will follow a simple
two-cutoff scaling scheme,\cite{Prigodin79,Firsov85,Bourbon91} in which  the flows of $V_\mu$
(Eqn.~(\ref{flowrpa})) and
$g_\mu$ (Eqn.~(\ref{twoloop}))  are stopped at $\ell_{x^1}$ and their
values are used as boundary conditions for the outer shell integration below $E_0(\ell_{x^1})$  
which is done with respect  to
constant warped energy surfaces (Figure~\ref{Fsurface}).  The outer shell
integration will be done at the one-loop level in the so-called ladder approximation meaning   that
the interference between the Cooper and Peierls channels will be  neglected. This decoupling between the two
channels at $T\ll T_{x^1}$ is actually
 not a  bad approximation  for $T_{c,\mu}$ when the couplings are attractive and favor superconductivity or 
 when the nesting properties of the warped Fermi surface are good and a density-wave instability
is  expected.  The approximation becomes less justified, however, if for example  
nesting deviations are present in the spectrum and the instability of the Peierls channel is weakened. In
this case, it has been shown that the residual  interference between the two channels gives rise to
non-trivial influence on pairing which becomes non-uniform along the Fermi surface $-$ e.g. unconventional
superconductivity for repulsive interactions and anisotropy of the SDW gap.\cite{Duprat01}

\begin{figure} \centerline{\includegraphics[width=8cm]{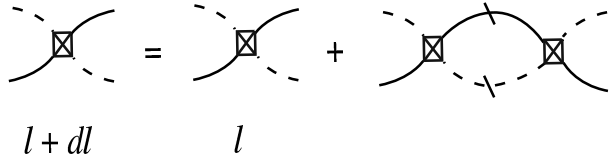}}
\caption{Recursion relation in the ladder approximation for the effective coupling ${\cal G}_\mu \equiv
V_\mu-g_\mu$ (crossed square) in the channel $\mu$ below the temperature of the single-particle
dimensionality crossover
$T_{x^1}$. }\label{Lad}\end{figure}

In the ladder approximation, the evaluation of $\dm\langle S^2_{\perp,2}\rangle_{\bar{0},c}$ at $\Qvec_0$ in
each channel  allows us to write
\begin{equation}
{dV_{\mu} \over d\ell}=  V_{\mu}g_{\mu} -
{1 \over 2} (V_{\mu})^2.
\end{equation}
This expression is incomplete, however, because  ${1\over 2}\langle S^{2}_{I,2} \rangle_{\0til,c}$
evaluated   at
${\bf Q}_0$, gives a logarithmic outer shell correction
of
the form ${1\over 2}g_{\mu}^2 d\ell$, which leads to the flow equation
\begin{equation}
{dg_{\mu} \over d\ell}=  
{1 \over 2} (g_{\mu})^2.
\end{equation}
In the action, the  corresponding combination of couplings is
proportional to
$O^*_{\mu}O_{\mu}$, and will  then add  to  $V_{\mu}$. Therefore it is useful to define ${\cal G}_{\mu}
\equiv V_{\mu} -g_{\mu}$ for which one obtains the flow equation of Fig.~\ref{Lad}  
\begin{equation}
{d \over d\ell}{\cal G}_{\mu}=  -{1\over 2} ({\cal G}_{\mu})^2.
\label{ladder}
\end{equation}
The relevant parameter
space for the action for $ E_0(\ell)< E_0(\ell_{x^1})$ in the ladder approximation becomes
\begin{equation}
\mu_S(\ell>\ell_{x^1})=(G^0_p(\kvec,\omega_n),{\cal G}_\mu). 
\end{equation}
The solution of (\ref{ladder})
has the following simple pole structure
\begin{equation}
{\cal G}_{\mu}(\ell)= {  {\cal G}_{\mu}(\ell_{x^1})
\over 1 + {1 \over 2} {\cal G}_{\mu}(\ell_{x^1})(\ell - \ell_{x^1})},
\label{ladderb}
\end{equation}
where ${\cal G}_{\mu}(\ell_{x^1})$ is the boundary condition 
deduced from the 1D scaling equations (\ref{flowrpa}) and
(\ref{twoloop}) at $\ell_{x^1}$. The presence of an instability of the Fermi liquid towards the condensation
of pairs in the
channel $\mu$ occurs at the  
critical temperature  
\begin{equation}
T_{c,\mu}=T_{x^1}\ \exp\bigl( -2 /
\mid {\cal G}_{\mu}(\ell_{x^1}) \mid\bigr).
\label{BCS} 
\end{equation}
 This BCS type of instability occurs when the effective  coupling constant  
 is  attractive (${\cal G}_\mu<0$). The phase diagram coincides 
with the one found in 1D when there is perfect
nesting (see Fig.~\ref{1DPhases}).  The 
critical temperature in each sector  is
associated with those correlations having the largest amplitude. The
variation of the critical temperature as a function of the interaction strength  is shown by the dashed line 
in Figure~\ref{Tcprofile}.  We also notice  that the BCS expression (\ref{BCS}) satisfies the extended
scaling hypothesis (\ref{scalingTc}), namely  $T_{c,\mu} \propto T_{x^1}\propto t_\perp^{1/\phi_{x^1}}$. 

\subsubsection{Response functions in the deconfined region}
By adding the coupling of fermion pairs to source fields, the calculation of 
response functions in the channel
$\mu$ below $T_{x^1}$ can readily be done. In the ladder approximation, the outer shell contraction 
$\dm\langle S_{h}S_{I,2} \rangle_{\bar{0},c}$, which amounts making the substitution $g_\mu \to {\cal
G}_\mu$ in Fig.~\ref{ChiRG}, leads at once to the flow equation  for the pair vertex part
\begin{equation}
{d\over d \ell} \ln z_\mu^h = -\dm {\cal G}_\mu.
\end{equation}
 Using the ladder result (\ref{ladderb}), one gets for the auxiliary susceptibility
$\bar{\chi}_\mu=(z_\mu^h)^2$ at
$\ell\ge\ell_{x^1}$
\begin{equation}
\bar{\chi}_\mu(\Qvec_0,\ell) = {\bar{\chi}_\mu(\Qvec_0,\ell_{x^1})\over [1 +\dm {\cal
G}_\mu(\ell_{x^1})(\ell-\ell_{x^1})]^2},
\end{equation}
which presents a power law singularity that is quadratic in $(T-T_{c,\mu})^{-1}$ as the temperature
approaches $T_{c,\mu}$ from above. The integration of this expression following the definition given in
Eqn.~(\ref{RGresponse}) causes no difficulties and one finds  for the response function
\begin{equation}
\chi_\mu(\Qvec_0,\ell) =   \chi_\mu(\Qvec_0,\ell_{x^1}) -(\pi v_F)^{-1} {\chi_\mu(\Qvec_0,\ell_{x^1})
(\ell-\ell_{x^1})\over 1+ \dm {\cal G}_\mu(\ell_{x^1})(\ell -\ell_{x^1})}.
\end{equation}
The singularity appears in the second term  and is    linear in $(T-T_{c,\mu})^{-1}$, which is typical of
a  ladder approximation. The critical exponent defined in Eqn.~(\ref{chitc}) is $\dot{\gamma}=1$  and  
corresponds to the  prediction of  mean-field theory. As for the amplitude of the singular behavior, it 
satisfies the extended scaling constraint shown in (\ref{constraint}). It is worth noting that these scaling
properties are also satisfied by the auxiliary susceptibility.

In going beyond the ladder approximation, the RG technique for  fermions
becomes less reliable in the treatment  of fluctuation effects beyond the Gaussian level which are usually
found sufficiently close the critical point $-$ at least in three dimensions. 
Standard approaches of critical phenomena, such as the statistical mechanics of the order
parameter from a  Ginzburg-Landau free energy functional, can be used  sufficiently close to
$T_{c,\mu}$.

\section{Kohn-Luttinger mechanism in quasi-one-dimensional metals}
\subsection{Generation of interchain pairing channels}
\begin{figure} \centerline{\includegraphics[width=12cm]{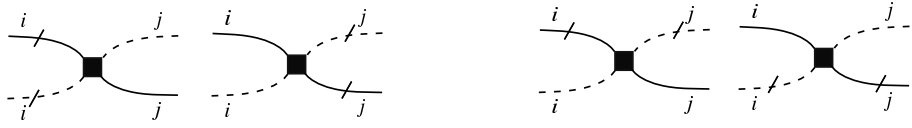}}
\caption{ Possible  outer-shell contractions for interchain pair hopping: intrachain channel
$\mu$ (left) and interchain channel $\bar{\mu}$ (right). The black square corresponds
to the pair hopping amplitude
$V_{\mu ij}$ between chains
$i$ and
$j$. }\label{pairhop}\end{figure}

The outer shell partial integration described in the RPA treatment of \S~\ref{Spairhopping} for $T>T_{x^1}$,
namely   in the presence of interchain pair hopping,  was restricted to the channel $\mu$ corresponding to
the most singular correlations of the 1D problem (Figure~\ref{1DPhases}). However, looking more closely  at
the possible contractions of fermion fields in $S_{\perp,2},$ a new   pairing possibility  emerges as  shown
in Figure~\ref{pairhop}. This consists of putting the two outer-shell fermion fields on separate chains
$i$ and
$j$, which can then be equated  with {\it interchain pairing}. The attraction between the two particles in 
the pair is mediated by the combination of pair hopping and  intrachain
correlations. Thus interchain pairing is  similar to  the Kohn-Luttinger mechanism for superconductivity
 by which Friedel charge oscillations can lead to an effective attraction between two electrons  and in turn
to an instability of the 3D metal at a finite but extremely low temperature.\cite{Kohn65} In quasi-1D
systems, however,  the K-L mechanism has a dual nature given that interchain pair hopping can exist not only 
in  the  Peierls channel but also in the  Cooper channel. Therefore for a Josephson coupling, the
interchain outer-shell decomposition  of Figure~\ref{pairhop} leads to an attraction between a particle and
hole on different chains. In this  case the attraction is mediated by the exchange of
Cooper pairs.   
\subsubsection{Coupling  constants}
In the following we will  restrict our description of the onset of interchain pairing correlations 
within the KW RG to the incommensurate case where only $g_1$ and $g_2$ scattering processes are present.
Thus in the temperature domain above $T_{x^1}$, which corresponds to $\ell <\ell_{x^1}$, we have seen in
\S~\ref{Spairhopping} that the outershell contraction in which the fermion lines are diagonal in the chain
indices yields the RPA equation  (\ref{flowrpa}). As pointed out above, another possible  
contraction would be to put the two outer shell fermion fields on distinct chains.

 This is best  illustrated
by first rewriting the pair hopping term (\ref{Vmu}) in terms of chains indices. Then putting   two
fields in the outershell, we consider the sum of two contributions
\begin{eqnarray}
  {S}_{\perp,2}  + {S'}_{\perp,2} =
  & -&\, {1\over 2}\pi v_F \, z^{-2}(\ell)\sum_{\mu,\tilde{q}}\sum_{i,j} V_{\mu,
ij}(\ell)\bigl(\bar{O}_{\mu,i}^*(\tilde{q})O_{\mu,j}(\tilde{q}) +
{\rm c.c} \ \bigr)\cr
                    &  + &\,{1\over 2} \pi
v_F \, z^{-2}(\ell)\sum_{\bar{\mu},\tilde{q}}\sum_{i,j} U_{\bar{\mu},
ij}(\ell)\bigl(\bar{O}_{\bar{\mu},ij}^*(\tilde{q})O_{\bar{\mu},ji}(\tilde{q})
+ {\rm c.c} \
\bigr),
\end{eqnarray}
the first of which leads to (\ref{flowrpa}) at the one-loop level, while the
second   is connected to the interchain channel denoted  $\bar{\mu}$.
The interchain composite fields
are defined by
\begin{eqnarray}
O_{\bar{\mu}_C, ij}(\tilde{q}) =  (T/L)^{1/2} \ \sum_{\tilde{ k}}
\alpha\ \psi_{-,-\alpha,i}(-\tilde{
k})\sigma_{\bar{\mu}_C}^{\alpha\beta}\psi_{+,\beta,j}(\tilde{ k} + \tilde{q}) 
\end{eqnarray}
for  the singlet (ISS: ${\bar{\mu}_C}=0$)  and triplet (ITS:
${\bar{\mu}_C}=1,2,3$) 
    interchain Cooper channel $\bar{\mu}_C$, and by
\begin{eqnarray}
O_{\bar{\mu}_P,ij}(\tilde{q})=(T/L)^{1/2} \ \sum_{\tilde{
k}}\psi^*_{-,\alpha,i}(\tilde{ k} -\tilde{
q})\sigma_{\bar{\mu}_P}^{\alpha\beta}\psi_{+,\beta,j}(\tilde{ k}) 
\end{eqnarray}
   for the charge-density-wave (ICDW: ${\bar{\mu}_P}=0$) and
spin-density-wave (ISDW:
${\bar{\mu}_P}=1,2,3$) pair fields of the  interchain Peierls
channel $\bar{\mu}_P$. This
 introduces the combinations of couplings
\begin{equation}
U_{\bar{\mu},ij} = \sum_\mu c_\mu^{\bar{\mu}} V_{\mu, ij}.
\label{combina}
\end{equation}
The constants $c_\mu^{\bar{\mu}}$ are obtained from 
\begin{eqnarray}
U_{\bar{\mu}_{C(P)}=0,ij}&=  &\ {3\over 2} V_{\mu_{P(C)}\ne 0, ij}
-{1\over 2} V_{\mu_{P(C)}=0,ij}
   \hskip 1 truecm {\rm ISS}\ {\rm (ICDW)}\cr
   U_{\bar{\mu}_{C(P)}\ne 0,ij}&= &\ {1\over 2} V_{\mu_{P(C)}\ne 0, ij}  +
{1\over 2} V_{\mu_{P(C)}=0,ij} \hskip 1 truecm
{\rm ITS}\ {\rm (ISDW)}.
\label{combinb}
\end{eqnarray}
  Therefore at the one-loop level  
  in addition to the  RPA term, one obtains 
 the  term
\begin{eqnarray}
{1\over 2}\langle
({S}'_{\perp,2})^2\rangle_{\bar{0},c}= \ \pi v_F z^{-2}(\ell)
{d\ell\over 4}\sum_{\bar{\mu},\tilde{q}}\sum_{i,j}  U^2_{\bar{\mu},
ij}(\ell)\ O_{\bar{\mu},ij}^*(\tilde{q})O_{\bar{\mu},ij}(\tilde{q}),
\label{KLgenerators}
\end{eqnarray}
which is also logarithmic. Owing to the dependence on the chain indices of the last composite
field on the right hand side
of this equation, this term  cannot be rewritten in terms of the intrachain channels and therefore stands as
a  new coupling which we will write $\delta{\cal S}_\perp$. It must be added to the parameter space of
the action.  To do so, it
is convenient to recast (\ref{KLgenerators}) in terms of  interchain
backward and forward
scattering amplitudes:
\begin{eqnarray}
\delta{\cal S}_\perp =   - {T\over L}\pi v_F z^{-2}(\ell)\
d\ell\sum_{\{\alpha\},\tilde{q}}\sum_{i,j}
&&\ \bigl( f_{2,ij}(\ell)\delta_{\alpha_1\alpha_4}\delta_{\alpha_2\alpha_3}-
f_{1,ij}(\ell)\delta_{\alpha_1\alpha_3}\delta_{\alpha_2\alpha_4}\bigr)\cr
\times &&  \  \psi^*_{+,\alpha_1,i}(\tilde{k}_1 + \tilde{q})
\psi^*_{-,\alpha_2,j}(\tilde{k}_2 - \tilde{q})\psi_{-,\alpha_3,j}(\tilde{k}_2
)\psi_{+,\alpha_4,i}(\tilde{k}_1 ).
\label{fgenerator}
\end{eqnarray}

\begin{figure} \centerline{\includegraphics[width=7cm]{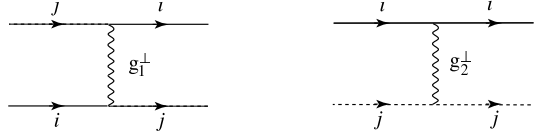}}
\caption{Perpendicular backward ($g_1^\perp$) and and forward ($
g_2^\perp$) scattering
terms generated by interchain pair tunneling.}
\label{couplings}\end{figure}

The outer-shell backscattering  amplitude $f_{1,ij}$   then
corresponds to intrachain momentum
transfer near
$2k_F$ between the particles  and a  change in the chain indices
during the process, whereas the
forward generating amplitude $f_{2,ij}$ corresponds to small intrachain momentum
with  no change in
the chain index (Figure~\ref{couplings}). The explicit expressions are 
\begin{eqnarray}
f_{1,ij}&= &-{1\over2} [V^2_{\mu_{P}\ne 0, ij}- V_{\mu_{P}\ne 0,
ij}V_{\mu_{P}= 0, ij}]
                 +{1\over2} [V^2_{\mu_{C}\ne 0, ij}- V_{\mu_{C}\ne 0,
ij}V_{\mu_{C}= 0, ij}]\cr\cr
2f_{2,ij}-f_{1,ij}
          &  =  & \ \  {1\over4} [3V^2_{\mu_{C}\ne 0, ij}+V^2_{\mu_{C}=
0, ij} ] +
{1\over4}[3V^2_{\mu_{C}\ne 0, ij}+V^2_{\mu_{P}= 0, ij}].
\label{generateurs}
\end{eqnarray}
Since the pair hopping amplitudes $V_\mu$ are themselves generated as a function of $\ell$, it is
clear that the $f's$ will be very small except when the $V_\mu$ approach the strong
coupling regime.     Successive KW transformations  will  lead to  the renormalization of the above
interchain backward and  forward scattering amplitudes  which we denote $g_{1,ij}^\perp $ and $g_{2,ij}^\perp
$, respectively. These define  the part of the action denoted ${\cal S}_\perp$, which  when expressed in
terms  of  uniform charge and spin  fields introduced in (\ref{uniform}), takes the rotationally
invariant form
\begin{eqnarray}
{\cal S}_\perp[\psi^*,\psi]_\ell=  -\pi
v_F z^{-2}(\ell)\sum_{p,\tilde{q}}\sum_{ij}\  \Bigl(\
(2g^\perp_{2,ij}(\ell)-g^\perp_{1,ij}(\ell))
\
\rho_{p,i}(\tilde{q})\rho_{-p,j}(-\tilde{q}) - \  g^\perp_{1,ij}(\ell)
\ {\bf S}_{p,i}(\tilde{q})\cdot {\bf S}_{-p,j}(-\tilde{q}) \ \Bigr)
\label{spincharge}
\end{eqnarray}
This expression clearly shows that interchain pair hopping generates a coupling between uniform charge and
spin excitations of different chains.   The flow equations of these couplings 
will now be  obtained  from ${1\over 2}\langle
{\cal S}_{\perp,2}^2\rangle_{\bar{0},c}$
at the one-loop
level. Aside from the  scale dependent generating terms, the contributions  consist of   
 the same type
of interfering Cooper and Peierls diagrams found in the  purely
one-dimensional problem (Fig.~\ref{recursion}). These outer shell logarithmic contributions can be
evaluated at once resulting in the following flow equations for the interchain spin and charge variables 
\begin{eqnarray}
{d\over d\ell} g_{1,ij}^\perp&=&f_{1,ij} -
(g_{1,ij}^\perp)^2\cr
{d\over d\ell}2g_{2,ij}^\perp-g_{1,ij}^\perp &= & 2f_{2,ij}^\perp-f_{1,ij}^\perp.
\label{flowgperp}
\end{eqnarray}
It is interesting to note
that, following the example of the purely one-dimensional couplings, the
interference between the
interchain and Cooper channels causes these equations to be decoupled.
Hence
 the spin and charge degrees of freedom  on 
different chains are not mixed through $g_1^\perp$ and $g_2^\perp$.

The solution for the charge coupling  is straightforward and  one obtains a
parquet-type solution
for  the spin part
\begin{equation}
g_{1,ij}^\perp(\ell)= { \bar{g}_{1,ij}^\perp(\ell)\over 1+
\bar{g}_{1,ij}^\perp(\ell)\ell}\ ,
\label{parquet}
\end{equation}
where
\begin{equation}
\bar{g}_{1,ij}^\perp(\ell)= \ \int_0^\ell d\ell' f_{1,ij}(\ell')[1 +
\bar{g}^\perp_{1,ij}(\ell')]^2.
\end{equation}
According to  (\ref{generateurs}),  $f_{1,ij}(\ell')$  and therefore
$\bar{g}_{1,ij}^\perp(\ell)$ are negative in the repulsive sector
($g_1>0$ and $2g_2-g_1 >
0$) where SDW correlations are singular (Figure~\ref{1DPhases}); 
   $g_{1,ij}^\perp(\ell)$ has then the possibility to flow to the strong attractive
coupling sector indicating the existence of  a spin
gap. Given the place held by $g_{1,ij}^\perp(\ell)$  in the hierarchy of generated couplings of the RG
transformation, however, this can only occur if  the
the SDW pair tunneling or exchange term $V_{\mu_P\ne 0,ij}$ first reaches the
 strong coupling domain $-$ a possibility  realized as the system enters  the Gaussian critical domain
near $T_{x^2}$. Since the above flow equations (\ref{flowgperp})  hold  for all pairs of chains $i$$j$
for $\ell<\ell_{x^1}$, the amplitude of the spin gap goes to zero as the separation $|i-j|\to \infty$.
It is worth  noting that the  spin gap for nearest-neighbor  chains is reminiscent of the one occurring
in  the two-chain problem.\cite{Schulz96,Lin98}

\subsubsection{Interchain response functions}
To establish the nature of  
   correlations  introduced by interchain  pairing, it is convenient to express   outershell decomposition
of $S_{\perp} + {\cal S}_{\perp}$  in the corresponding channel 
\begin{eqnarray}
S_{\perp,2} + {\cal S}_{\perp,2}\,\Big|_{\bar{\mu}^\pm} =  { 1\over 2}\pi v_F z^{-2}  
\sum_{\tilde{q},\bar{\mu}^\pm}\sum_{ij} \  
   g^\perp_{\bar{\mu}^\pm,ij}(\ell) \bar{{\cal O}}^{
*}_{\bar{\mu}^\pm,
ij}(\tilde{q}) {\cal O}_{\bar{\mu}^\pm, ij}(\tilde{q})  + {\rm c.c}, 
\label{shellinter}
\end{eqnarray}
where we have introduced the combinations of couplings
\begin{eqnarray}
 g^\perp_{{\rm ISS}^\pm,ij}& =  & -g^\perp_{1,ij} - g^\perp_{2,ij} \pm  U_{{\rm ISS},ij} \cr
g^\perp_{{\rm ITS}^\pm,ij} &=   & \ g^\perp_{1,ij} - g^\perp_{2,ij} \pm U_{{\rm ITS},ij} 
\label{gintera}
\end{eqnarray}
for symmetric $(+)$  and antisymmetric $(-)$ interchain superconductivity and their analog expressions
\begin{eqnarray}
 g^\perp_{{\rm ICDW}^\pm,ij} &=   & g^\perp_{2,ij} - 2g^\perp_{1,ij} \pm  U_{{\rm ICDW},ij} \cr
g^\perp_{{\rm ISDW}^\pm,ij}& = & g^\perp_{2,ij} \pm U_{{\rm ISDW},ij} 
\end{eqnarray}
for interchain density-wave. The corresponding pair fields  are given by 
\begin{equation}
{\cal O}_{\bar{\mu}^\pm, ij}= {1\over 2} \bigl(O_{\bar{\mu}, ij} \pm O_{\bar{\mu}, ji} \bigr)
\end{equation}
for  symmetric $(+)$ and antisymmetric $(-)$ composite fields. 

\begin{figure} \centerline{\includegraphics[width=7cm]{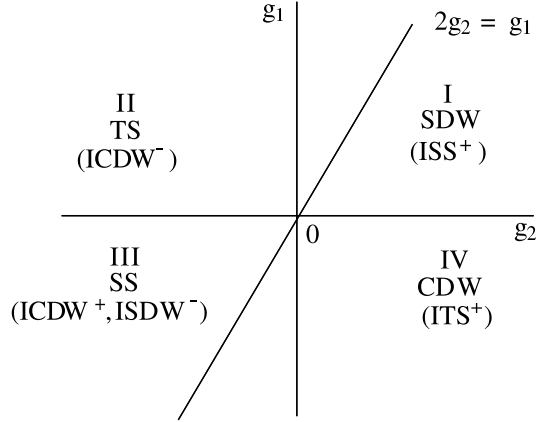}}
\caption{Phase diagram of the quasi-one-dimensional electron gas with
interchain hopping and no Umklapp scattering.
Dominant  correlations of intrachain
(interchain) pairing channels are
given.}\label{Phaseinter}\end{figure}

We add  the  coupling to  source fields
\hbox{$ S_h= \sum O^*_{\bar{\mu}^\pm,ij}h_{\bar{\mu}^\pm,ij} + {\rm c.c.}$}.
The one-loop corrections to the pair vertex part are obtained from $\langle (S_{\perp,2} + {\cal
S}_{\perp,2}){S}_{h}\rangle_{\bar{0},c}$ (which is equivalent to making the substitution $g_ \mu
\to g^\perp_{{\rm ISS}^\pm,ij}$ in Fig.~\ref{ChiRG}), and yields the flow equations pair vertex
renormalization factor
$z_{\bar{\mu}^\pm,ij}$ and hence  that of the auxiliary susceptibility $\bar{\chi}_{\bar{\mu}^\pm,ij}=
z_{\bar{\mu}^\pm,ij}^2$ (with $z=1$ ):
\begin{equation}
{d\over d\ell}\ln \bar{\chi}_{\bar{\mu}^\pm,ij}=
g^\perp_{\bar{\mu}^\pm,ij}.
\label{chibarinter}
\end{equation}
Obtaining an   expression for
$\bar{\chi}_{\bar{\mu}^\pm,ij}$ in closed form for
arbitrary chains $i$ and $j$ is not possible. However, in the case
where intrachain channels have
developed only short-range order and $U_{\bar{\mu},ij}$ is
only sizable for nearest-neighbour
chains, one can extract the dominant trend of interchain  auxiliary
suceptibilities in the various
sectors of the $g_1g_2$ plane. Thus by taking in Fourier space
$V_\mu(\Qvec,\ell)\approx
(-1)^{\delta_{\mu\mu_c}}\mid V_{\mu}(\ell)\mid \cos q_\perp $
for  the pair tunneling term
(\ref{VRPA}), one gets
$V_{\mu,ij}(\ell) = {1\over 2}(-1)^{\delta_{\mu\mu_c}} \mid
V_{\mu}(\ell)\mid $ where $\mid
V_{\mu}(\ell)\mid =\mid V_{\mu}(\Qvec_0,\ell)\mid $ is the maximum
amplitude  of
$V_\mu(\Qvec,\ell)$. This approximation  holds sufficiently far from $T_{c,\mu}$, namely for 
${1\over 2} \bar{V}_\mu\chi_\mu/\bar{\chi}_\mu$ not too close to unity. The
nearest-neighbour
$ij$  auxiliary susceptibility  then becomes
\begin{eqnarray}
\bar{\chi}_{\bar{\mu}^\pm, ij}(\ell) &=   &
X_{\bar{\mu},
ij}(\ell)\exp\left\{\pm {1\over 2}\sum_{\mu}\int _0^\ell
c_\mu^{\bar{\mu}} \mid
V_{\mu}(\ell')\mid d\ell' \right\} \cr
&\approx  &   X_{\bar{\mu}, ij}(\ell) \prod_\mu
\Bigl[1-{1\over 2} (\bar{\chi}_\mu(\ell))^{-1}
\bar{V}_\mu(\ell)\chi_\mu(\ell)\Bigr]^{\mp c_\mu^{\bar{\mu}}},
\end{eqnarray}
where the values of the channel index $\mu $ and the exponent
$c_\mu^{\bar{\mu}}$ are  given
in (\ref{combina}-\ref{combinb}). The  transient amplitude is given by 
\begin{eqnarray}
      X_{\bar{\mu}, ij}(\ell) =     \ \exp\left\{\int _0^\ell
   \beta_{\bar{\mu}} d\ell' \right\}, 
\end{eqnarray}
where for superconductivity $\beta_{\rm ISS}=-g^\perp_{1,ij} - g^\perp_{2,ij} $  and
$\beta_{\rm ITS}=g^\perp_{1,ij} - g^\perp_{2,ij} $, while for density-wave  $\beta_{\rm
ICDW}=g^\perp_{2,ij} - 2g^\perp_{1,ij}
$ and $\beta_{\rm ISDW}=g^\perp_{2,ij} $.  The onset of  interchain
correlations, even though they are small,  will  be essentially  determined   by the most singular
correlations of the intrachain channels $\mu$ (Figure~\ref{Phaseinter}) and the exponent
$c_\mu^{\bar{\mu}}$. At finite temperature, we can write
\begin{equation}
\bar{\chi}_{\bar{\mu}^\pm, ij}(T)\approx X_{\bar{\mu}, ij}(T) \prod_\mu \ \Bigl[1-{1\over 2}
(\bar{\chi}_\mu(T))^{-1}
\bar{V}_\mu(T)\chi_\mu(T)\Bigr]^{\mp c_\mu^{\bar{\mu}}}.
\end{equation}
In the repulsive sector where
$g_1> 0, 2g_2-g_1> 0$,   SDW correlations are singular (region
I, Figure~\ref{Phaseinter}) and   lead to  an enhancement  of
symmetric (+) interchain
singlet superconducting (ISS$^+$) correlations with $c_{\mu_P\ne
0}^{\bar{\mu}_C=0}= 3/2 $. It should
be noted here  that, in this sector,   symmetric interchain  triplet
(ITS$^+$)  correlations are also
enhanced by SDW correlations with $c_{\mu_P\ne 0}^{\bar{\mu}_C\ne 0}= 1/2 $. The pairing on different chains
results from the exchange of spin fluctuations.  On the T-L line at
$g_1=0$ and
$g_2> 0$,  SDW and CDW correlations are equally singular leading to 
an  enhancement  of both ISS$^+$ and ITS$^+$  
correlations with a net exponent of unity.

 In  region II of Figure
\ref{Phaseinter}, where  $g_1>0$,
$2g_2-g_1< 0$, the TS channel is the most singularly enhanced  
leading to interchain charge-density-wave (ICDW$^-$) correlations with $ c_{\mu_P\ne
0}^{\bar{\mu}_P=0}= 3/2 $. Here  the pairing  results from  a
three-component (triplet) Cooper pair
exchange between the chains.  In the same region, an enhancement,
although smaller,
of  
antisymmetric interchain
spin-density-wave  correlations (ISDW$^-$) is also present with  $ c_{\mu_P\ne
0}^{\bar{\mu}_P\ne 0}= 1/2 $. On the T-L line at $g_2<0$,
however, SS and TS  are equally singular so
both contribute to give the
same enhancement for ISDW$^-$ and  ICDW$^-$. If one moves to  the attractive sector (region
III, Figure~\ref{Phaseinter}),   where $g_1<0$, $2g_2-g_1<0$ and SS is the most singular channel 
along the chains, one finds that
 symmetric  ICDW$^+$ and   antisymmetric  ISDW$^-$ channels  have
similar enhancements 
$ c_{\mu_C=
0}^{\bar{\mu}_P=0}=c_{\mu_C=
0}^{\bar{\mu}_P\ne 0}= 1/2$.
 Finally in the region IV, where $g_1<0$, $2g_2-g_1> 0$ and CDW correlations are the most singular,  
symmetric ITS$^+$ correlations are enhanced in the interchain triplet channel with the exponent $c_{\mu_P=
0}^{\bar{\mu}_C\ne 0} =1/2 $.

\subsection{Possibility of long-range order in the interchain pairing channels}
If the initial parameters of the model are such that the  system  can reach the energy or temperature 
domain  $T_{x^1} $ without developing long-range order, the pair-hopping amplitude $V_\mu$ is still in the
weak coupling regime. We have seen in \S \ref{BelowTx1}, however,   that infrared singularities in both the
Peierls and Cooper channels persist above $\ell_{x^1}$ so that long-range order  in the   intrachain channel
$\mu$  can occur if  
${\cal G}_\mu <0$. This will be so in the Peierls channel when  deviations from perfect
nesting  of the warped Fermi surface do not exceed a given threshold  (Fig.~\ref{Fsurface}-a), while in the
Cooper channel if  a  magnetic field  is  below  a critical value. Therefore sufficiently large nesting
deviations or magnetic field must be introduced in the model in order for long-range order in the
  {\it interchain}  channels  to be  possible. We will now illustrate how such a possibility can be
realized   in the  repulsive case when   nesting deviations are introduced in the model. We will  thus
focus our attention in the region I of the phase diagram where symmetric interchain singlet superconductivity
(ISS$^+$) can be  stabilized (Fig.~\ref{Phaseinter}).  
A similar description holds for the Peierls interchain channel in the presence of a sufficiently strong 
magnetic field (regions II and III in Fig.~\ref{Phaseinter}).

 Frustration of nesting is commonly introduced  by considering the possibility of particles
 hopping  
 to second nearest-neighbor chains, i.e. by adding the term 
$-2t_{\perp2}\cos 2k_\perp
$  to the spectrum at $\ell=0$,  where $t_{\perp2}$ is the amplitude of the hopping ($t_{\perp2}\ll
t_\perp$).  The relevant parameter space of the action  at the crossover  value
$\ell_{x^1}$ becomes 
$\mu_S(\ell_{x^1})= (G_p ^0(\kvec,\omega_n),{\cal G}_\mu,g^\perp_{\bar{\mu}^\pm})$, where
\begin{equation}
G_p^0(\kvec,\omega_n)= {z\over i\omega_n -\epsilon_p(k) + 2t^*_\perp\cos k_\perp + 2t^*_{\perp2}\cos
2k_\perp}
\end{equation}
is  the effective propagator in which $z$ is the quasi-particle weight at $\ell_{x^1}$ and 
$t_{\perp(2)}^*\equiv zt_{\perp(2)}$. 
 If one cranks up the  amplitude of $t_{\perp 2}^*$,  nesting  of the
Fermi surface at
$\Qvec_0=(2k_F,\pi)$ is quickly  suppressed so that  the $T_{\rm c,SDW}$ in  the region I of the phase
diagram 
  decreases rapidly and eventually vanishes above the critical $t^{*c}_{\perp 2}$\cite{Yamaji82}. As for the
infrared singularity of the  Cooper channel, however, it is still present and can  then lead to an
instability of the normal state. 

We will adopt here a simple ladder RG description of this instability,
which neglects the interference between the $-$ weakened $-$ Peierls channel and  the Cooper one. This
approximation should be qualitatively valid for   $t_{\perp 2}^*\gg t^{*c}_{\perp 2}$ (a more quantitative
description including the interference effect can be found in Ref.~\cite{Duprat01}). We will further assume
for simplicity that the enhancement of ${\cal G}_\mu$ above $\ell_{x^1 }$ is negligible and the  
transverse short-range order of the chains is entirely contained in the $ij$ dependence of
$g^\perp_{{\rm ISS}^+,ij}$  in Eqns.~(\ref{gintera}) and (\ref{combinb}), which is restricted to 
nearest-neighbor chains.  In this way, the repulsive coupling ${\cal G}_\mu$ is entirely local and does not
compete with the interchain part: fermions on different chains  are said to  avoid 
intrachain repulsion.\cite{Emery86,BealMonod86}  In transverse Fourier space, the
outer shell decomposition (\ref{shellinter}) in the channel ISS$^+$ becomes
\begin{eqnarray}
S_{\perp,2} + {\cal S}_{\perp,2}\,\Big|_{{\rm ISS}^+} =  { 1\over 2}\pi v_F z^{-2}  
\sum_{\tilde{q}}\sum_{k_\perp,k'_\perp} \  
       g^\perp_{{\rm SS}}(\ell)\,{\eta_{k_\perp}} \bar{O}^{
*}_{{\rm SS}}(\tilde{q},k_\perp) {\eta_{k'_\perp}}{ O}_{{\rm SS}}(\tilde{q},k'_\perp)  + {\rm
c.c}
\label{ISS} 
\end{eqnarray}
in which  we have defined ${ O}_{{\rm SS}}(\tilde{q})= \sum_{k_\perp}{ O}_{{\rm
SS}}(\tilde{q},k_\perp)$ and where $\eta_{k_\perp}=
\cos k_\perp$ is the 
Fourier factor that  leads to a global antisymmetric order parameter for interchain singlet
superconductivity. The combination of couplings at $\ell$  
\begin{equation}
g^\perp_{{\rm ISS}^+}(\ell)\equiv -g^\perp_{1}(\ell)
- g^\perp_{2}(\ell) +  U_{{\rm ISS}}(\ell)
\end{equation}
is obtained from the amplitude of the transverse couplings at $\ell$.  In the ladder approximation, the
outer shell integration
$\dm
\langle (S_{\perp,2} + {\cal S}_{\perp,2})^2\rangle_{\bar{0},c}$  is done in this channel. At zero
external pair momentum, this leads to the flow equation 
\begin{equation}
{d\over d\ell} g^\perp_{{\rm ISS}^+}=  {1\over4} (g^\perp_{{\rm ISS}^+})^2,
\end{equation}
which is  easily integrated to give the simple pole expression
\begin{equation}
 g^\perp_{{\rm ISS}^+}(\ell) =  {g^\perp_{{\rm ISS}^+}(\ell_{x^1})\over 1-  {1\over4}g^\perp_{{\rm
ISS}^+}(\ell_{x^1})(\ell-\ell_{x^1})}.
\label{ginter}
\end{equation}
The critical temperature  for the superconducting instability is then  $T_c=T_{x^1} \exp[-4/g^\perp_{{\rm
ISS}^+}(\ell_{x^1})]$ where $ g^\perp_{{\rm
ISS}^+}(\ell_{x^1})>0$ corresponds to a net attraction in the interchain superconducting channel. From
(\ref{ISS}), the singlet order parameter is of the form
\begin{equation}
\Delta(k_\perp) = \Delta_0\eta_{k_\perp},
\end{equation}
which has nodes on the Fermi surface.   

The calculation of the interchain superconducting response  causes no particular difficulty. In effect, by
adding  the coupling to a source field (\hbox{$ S_h= \sum
{{O}}^*_{\bar{\mu}^+}(\qtil,k_\perp)\eta_{k_\perp}h_{\bar{\mu}^+} + {\rm c.c.}$}), one obtains 
the flow equation for the auxiliary susceptibility
\begin{equation}
{d\over d\ell} \ln \bar{\chi}_{{\rm ISS}^+} =  \dm g^\perp_{{\rm ISS}^+}.
\end{equation}
Using  (\ref{ginter}), this  is integrated to give the double pole singularity
\begin{equation}
\bar{\chi}_{{\rm ISS}^+}(\ell) = {\bar{\chi}_{{\rm ISS}^+}(\ell_{x^1})\over [ 1- {1\over 4} g^\perp_{{\rm
ISS}^+}(\ell_{x^1})(\ell-\ell_{x^1})]^2}, 
\end{equation}
where the boundary condition $\bar{\chi}_{{\rm ISS}^+}(\ell_{x^1}) $ is given by the solution of
(\ref{chibarinter}) at $\ell_{x^1}$. The loop integration of the above expression following the definition 
(\ref{RGresponse}), leads to  the response function
\begin{equation}
\chi_{{\rm ISS}^+}(\ell)=\chi_{{\rm ISS}^+}(\ell_{x^1}) -(2\pi v_F)^{-1} {\bar{\chi}_{{\rm
ISS}^+}(\ell_{x^1}) (\ell-\ell_{x^1})\over 1-{1\over 4} g^\perp_{{\rm
ISS}^+}(\ell_{x^1})(\ell-\ell_{x^1})}
\end{equation}
 which has the expected simple pole structure  of  the ladder approximation.

\section{Summary and concluding remarks}  
  We  have  reexamined    the  scaling tools that enter  in the description
of quasi-one-dimensional interacting fermion systems. On phenomenological  grounds, we have emphasized that
the scaling ansatz can provide a qualitative understanding of the existence of power law  behavior of various
quantities in one dimension while its
 extension for non zero interchain coupling  has allowed us to describe the different forms of dimensionality
crossover   in the quasi-one-dimensional case.  A key feature of this phenomenology is the
existence of fluctuations of single fermion and pair degrees of freedom on many length or energy
scales, the origin of which 
  can be microscopically understood  using the renormalization group language.  We found it useful to 
reconsider the formulation of this methodology and we have  seen that the adoption of a classical Wilson
procedure for the Kadanoff transformation has a molding influence on the RG flow   beyond the one-loop level.
Thus   the constraint of a sharp natural cutoff 
$k_0$   on the fermion spectrum which keeps the     loop momenta
 in the  
 outer energy shells  inevitably leads   to a non-local flow of the parameters that
define the action.  Non-locality   results from the inclusion   of
many-particle  marginal  interactions in the outer shell integration. These are not
present in the bare action but  are   generated along the RG flow.  The resulting structure of outer shell
integration finds an interesting parallel with the   functional description of
the renormalization group equations proposed by Polchinski in the framework of  the $\phi^4$
theory,\cite{Polchinski84} and used   by Zanchi and Schulz in the context of  two-dimensional interacting
fermion systems.\cite{Zanchi00}

 It is only when the marginally relevant or
irrelevant couplings are treated in the local approximation that purely logarithmic
scaling is  recovered. Although this has only been 
verified at the two-loop loop level and in the one-dimensional case, similar conclusions 
are expected to hold at higher order and in higher dimensions as well. The procedure described in this review
 casts    light  on a previous alternative  formulation of the  RG which is   known to offer a
more direct route to logarithmic scaling at high order  but which on the other hand requires the 
relaxation of some of the standard rules of the  
  Wilson method.  It also highlights  how the difficulties linked to non-locality of the flow equations are 
essentially  hidden in older versions of the renormalization group.    

The RG method  is widely recognized as a   tool well suited to the study of the  influence of   interchain
hopping  on the stability of the  Luttinger and Luther-Emery  liquid states   in  
one dimension. We have reviewed the conditions under which the  coupling between stacks can make these
non-Fermi liquid states unstable.
 A  possible instability occurs when the Luttinger liquid yields to the formation of a Fermi liquid. The
recovery of quasi-particle excitations,  here called  
     a dimensionality crossover for the fermion quantum coherence,    squares well with the
picture provided by the extended scaling ansatz. A second instability  emerges   
when the kinematics of interchain single-particle  hopping is combined  with intrachain pair
correlations. 
  Effective pair hopping processes between chains then   form and their coupling to singular  correlations
along the chains  lead to  a distinct (two-particle) dimensionality crossover. This
  is physically meaningful 
   as long as the corresponding temperature scale is above the one predicted for  single particles.
   Such a situation is favored by cranking up interactions which on the one hand increases pair correlations
and  on the other hand reduces single particle coherence.   The temperature scale for two-particle
dimensionality crossover marks  the onset of long-range order which accords well with the extended scaling
ansatz. For sufficiently  weak  interactions, however, single-particle deconfinement occurs first and the
possibility for long-range order remains as an instability  of the Fermi liquid.  In this case, the ladder
level of the RG  gives a BCS character to the transition.

The RG method can also be useful in  studying how the Kohn-Luttinger mechanism for unconventional pairing
works in the quasi-one-dimensional case. We have seen that this mechanism opens up   channels of
correlation  which pair  particles    on  different stacks as a result of pair hopping processes
between chains. Since these are  present in both the Peierls and Cooper channels,  the
Kohn-Luttinger mechanism has a dual nature. Thus it was shown   not only  that the exchange of  density-wave 
correlations  can lead to interchain Cooper pairing but also that interchain Peierls pairing between a
particle and a hole is also possible when both particles exchange  Cooper pairs. Long-range order in these
unconventional channels can only be achieved  from the  Fermi liquid state in which enough nesting
deviations or magnetic field are  present  to frustrate  the formation of  long-range order
in the  the primary channels of correlation along the chains.  

The results presented in this review   by no means exhaust the range of possibilities supplied by the
RG method. We will end this article by giving two specific examples. The
first concerns  the issue of single-particle deconfinement that leads to the recovery of a Fermi liquid.  
This question is of great interest especially in connection with the   normal state properties
obeserved in concrete realizations of   quasi-one-dimensional materials like the organic
conductors.\cite{Bourbon99,Biermann01}  We have seen that the critical domain of the primary (Luttinger
liquid) fixed point in a quasi-one-dimensional system essentially covers the whole temperature range of
interest. This favors in turn a slow and gradual crossover to a Fermi liquid, which means in practice that
transitory effects are likely to be important in the description of these materials. Although
crossover scaling effects have been underestimated in the treatment given above, they are not beyond the reach
 of  the RG method.  

Another promising  avenue is the application of the RG method to interacting fermions on a
lattice.\cite{Dumoulin96} The Kadanoff transformation can actually be generalized to fermions with the full
tight binding spectrum. Transients to the continuum limit can then be considered in a quantitative way. In
the case of the fermion representation of the one-dimensional spin chain, for example, this has been shown
to be an  essential ingredient  in the  description of the thermodynamics.

\section{Acknowledgements}
C.B. thanks L. G. Caron, N. Dupuis, A.-M. Tremblay and D. S\'en\'echal for numerous discussions and R. Shankar and G. Kotliar for interesting comments about the RG method at high order. This work is supported by the Natural Sciences and Engineering Research Council of Canada (NSERC), le Fonds pour la Formation de Chercheurs et
l'Aide \`a la Recherche du Gouvernement du Qu\'ebec (FCAR), and by the `superconductivity program' of the
Institut Canadien de Recherches Avanc\'ees (CIAR).

\appendix
\section{One-particle self-energy at the two-loop level}
\subsection{Backward and forward scattering contributions}
\label{six}
Following the partial trace operation in the absence of umklapp
scattering, three-particle interactions are generated
from the outer shell contraction $\delta S_\lambda^{(j)}\equiv \dm\langle
S_{I,1}^2\rangle_{\bar{0},c}$. The corresponding
expression  reads
\begin{eqnarray}
    \delta S^{(j)}_\lambda =
  \ (\pi v_F)^2{T^2\over L^2}\
\sum_{\{\alpha,\alpha'\}}\sum_{\kktil_{1,2}, \kktil'_2,\qtil'}\sumslashD_{\qtil}
 \lambda^+_{\{\alpha,\alpha'\}}(\ell_{j-3}) G^0_+(\kktil_1+\qtil) &&\!\!\!\!\!\!\!
\psi^*_{+,\alpha'_1}(\kktil_1+
\qtil'+\qtil)\psi^*_{-,\alpha'_2}(\kktil'_2-
\qtil')\psi^*_{-,\alpha_2}(\kktil_2-
\qtil)\cr
&\times&
\psi_{-,\alpha'_3}(\kktil'_2)\psi_{-,\alpha_3}(\kktil_2)
\psi_{+,\alpha_4}(\kktil_1) \cr
+\ (\pi v_F)^2{T^2\over L^2}\
\sum_{\{\alpha,\alpha'\}}\sum_{\kktil_{1}, \kktil'_{1,2},\qtil'}\sumslashD_{\qtil}
\lambda^-_{\{\alpha,\alpha'\}}(\ell_{j-3})G^0_-(\kktil_2-\qtil) &&\!\!\!\!\!\!\!
\psi^*_{-,\alpha'_2}(\kktil'_2-
\qtil')\psi^*_{+,\alpha'_1}(\kktil'_1+
\qtil')\psi^*_{+,\alpha_1}(\kktil_1+
\qtil)\cr
&\times  &
\psi_{+,\alpha'_4}(\kktil' _1)\psi_{+,\alpha_4}(\kktil_1)\psi_{-,\alpha_3}(\kktil_2)
\label{lambda}
\end{eqnarray}
where  $\lambda^+_{\{\alpha,\alpha'\}}(\ell_{j-3}) =
g_{\{\alpha\}}(\ell_{j-3})g_{\{\alpha'\}}(\ell_{j-3})\delta_{\alpha_1\alpha'_4}$,
$\lambda^-_{\{\alpha,\alpha'\}}(\ell_{j-3}) =
g_{\{\alpha\}}(\ell_{j-3})g_{\{\alpha'\}}(\ell_{j-3})\delta_{\alpha_2\alpha'_3}$, $\ell_{j-3}=
\ln (E_0/E_0(\ell_{j-3}))$ and
$g_{\{\alpha\}}$ is given by (\ref{gincom}).  Here
the summation $\sumslash_{\qtil}$ covers all $\omega_m$ but $q$ is restricted such that
$k_1+q$ ($k_2-q)$ in the propagator $G_+^0$ ($G_-^0$) is in the outer
momentum  shell at the step $(j-2)d\ell
$ for $j\ge3$. The   two terms in (\ref{lambda}),  which depends on the step
$j$, correspond to the
diagrams of Figure \ref{sixth}-a. At the
next partial integration $(j-1)d\ell$ , the above contributions will  in
turn  be
contracted leading to a second fermion line at
the adjacent  momentum shell
  (Figure \ref{sixth}-b). When this set  of contractions is repeated  as a function of $j$, contributions add
to give an effective two-particle interaction of the form 
\begin{eqnarray}
 \delta S'_{\lambda}\equiv && \sum_{j=3}^{N-1}\langle \delta S^{(j)}_{\lambda,2}
\rangle_{\bar{0},c}\cr
= && (\pi v_F)^2 {T^2\over L^2}\sum_{\kktil,\qtil'}\sum_{j=3}^{N-1}\sum_{\{\alpha,\alpha'\}}
\sumslashD_{\kktil'}\!\!{}^{^*}\
\sumslashD_{\qtil}\lambda^+_{\{\alpha,\alpha'\}}(\ell_{j-3}) \Big\{
\delta_{\alpha'_3\alpha_2}\, G^0_+(\kktil +\qtil)G_-^0(\kktil'-\qtil) \cr
&& \hskip 5 truecm \times \,
\psi^*_{+,\alpha'_1}(\kktil + \qtil' +\qtil )\psi^*_{-,\alpha'_2}(\kktil'-\qtil'-\qtil)
  \psi_{-,\alpha_3}(\kktil')\psi_{+,\alpha_4}(\kktil)\cr
&& \hskip 6.0  truecm  + \, \delta_{\alpha'_2\alpha_3} \, G^0_+(\kktil
+\qtil)G_-^0(\kktil'-\qtil') \cr
&&\hskip 5 truecm \times \,
\psi^*_{+,\alpha'_1}(\kktil + \qtil' +\qtil)\psi^*_{-,\alpha_2}(\kktil'-\qtil'-\qtil )
  \psi_{-,\alpha'_3}(\kktil')\psi_{+,\alpha_4}(\kktil) \Big\}\cr
 + \ &&  (\pi v_F)^2 {T^2\over
L^2}\sum_{\{\alpha,\alpha'\}}\sum_{\kktil}\sum_{j=3}^{N-1}\sumslashD_{\kktil_1}\!\!{}^{}\
\sumslashD_{\qtil}\lambda^+_{\{\alpha,\alpha'\}}(\ell_{j-3}) \, \delta_{\alpha'_1,\alpha_4} \, G^0_+(\kktil_1
+\qtil)G^0_+(\kktil_1) \cr
&&\hskip  5 truecm \times \,
\psi^*_{-,\alpha'_2}(\kktil' + \qtil)\psi^*_{-,\alpha_2}(\kktil- \qtil)
  \psi_{-,\alpha'_3}(\kktil')\psi_{-,\alpha_3}(\kktil)
\label{Sprime}
\end{eqnarray}
for $\lambda^+$ and   an analogous expression  for $\lambda^-$.
In the first two terms,  the summation
$\sumslash_{\!\kktil'}^{\!*}$ covers all
fermion frequencies $\omega_{n'}$ and  its momentum interval is  such that
$k'-q$  is in the outer shell while $k'$ is still in
the inner shell.  These expressions  correspond  to the two series of diagrams shown in
Figure~\ref{sixth}-c,  while the last term (not shown in Figure~\ref{sixth}-c) only involves outgoing
fermions near $-k_F$.     The final contraction of the cascade is obtained by putting
$k'$ in the outer momentum shell at $Nd\ell$
\begin{eqnarray}
 \langle \delta S'_{\lambda,2} \rangle_{\bar{0},c}
\ = && 2(\pi v_F)^2 {T^2\over L^2}\sum_{\alpha}\sum_{j=3}^{N-1}\sumslashD_{\kktil'}\!\!{}^{}\
\sumslashD_{\qtil}\, (g_1^2(\ell_{j-3})+g_2^2(\ell_{j-3}) -g_1(\ell_{j-3})g_2(\ell_{j-3}))\cr
 &&  \times \, \Big\{\Big[G^0_+(\kktil+\qtil)G_-^0(\kktil'-\qtil)G_-^0(\kktil')
+G^0_+(\kktil+\qtil)G_-^0(\kktil'+\qtil)G_-^0(\kktil') \Big]\,
\psi^*_{+,\alpha}(\kktil)\psi_{+,\alpha}(\kktil)\cr
  &&+ \  \Big[G^0_+(\kktil'+\qtil)G_-^0(\kktil-\qtil)G_-^0(\kktil')
+\,  \ldots\, \Big]\psi^*_{-,\alpha}(\kktil)\psi_{-,\alpha}(\kktil)   \Big\}.
\label{Green}
\end{eqnarray}
The first two  terms in the sum lead to a self-energy correction for fermions (on the  $+k_F$ branch) the
momenta, $k$, of which are 
 located at the top of the inner shell. The outer shell contractions leading to the
last terms  are self-energy corrections for   fermions on the $-k_F$ branch (only one such term is
shown explicitly).  Owing to the outer shell constraints put on the momentum, however,  when the summations 
over  fermion and boson frequencies are carried out  in the low temperature limit,  all these $-k_F$
branch terms give  a vanishingly small contribution in comparison with the  first two terms and can  be
safely ignored.   
\begin{eqnarray}
\langle \delta S'_{\lambda,2} \rangle_{\bar{0},c}
  = && -\sum_{\alpha}
\left\{\int_{\dm E_0(\ell_N)}^{\dm E_0(\ell_{N-1})} \sum^{N-1}_{j=3}
\int_{-\dm E_0(\ell_N)- \dm E_0(\ell_{j-3 })}^{-{\dm E_0(\ell_N) -\dm E_0(\ell_{j-2})}}  +
\int_{-\dm E_0(\ell_{N-1})}^{-\dm E_0(\ell_{N})} \sum^{N-1}_{j=3}
\int_{\dm E_0(\ell_N) +\dm E_0(\ell_{j-2})}^{\dm E_0(\ell_N)+ \dm E_0(\ell_{j-3 })}
\right\} \, d\epsilon' d(v_Fq) \cr
&& \ \ \ \ \ \times\  (g_1^2(\ell_{j-3})+g_2^2(\ell_{j-3})
-g_1(\ell_{j-3})g_2(\ell_{j-3})) \ {1 \over
[G_+^0(\kktil)]^{-1} -2v_Fq } \,
\psi^*_{+\alpha}(\kktil)\psi_{+\alpha}(\kktil).
\label{nonlocal}
\end{eqnarray}
  The  integrand  turns out to be independent of
$\epsilon'\equiv\epsilon_-(k')$ and we will define $\delta E_0/2= \int^{-\dm E_0(\ell_N)}_{-\dm
E_0(\ell_{N-1})}d\epsilon' = \int_{\dm E_0(\ell_N)}^{\dm
E_0(\ell_{N-1})}d\epsilon'$. Now assuming that the coupling constants 
vary slowly as a function of
$j$, this expression can be evaluated in the local  approximation (Figure~\ref{sixth}-d):
\begin{eqnarray}
\langle \delta S'_{\lambda,2} \rangle_{\bar{0},c}&=&  \!\!\!\! -
{\delta E_0\over 2}\sum_{\alpha}\left\{\int_{-\dm(E_0(\ell)+E_0)}^{- E_0(\ell))}
\!\! +\int_{E_0(\ell)}^{\dm(E_0(\ell)+E_0)} \right\}  (g_1^2(\ell_q)+g_2^2(\ell_q)
-g_1(\ell_q)g_2(\ell_q)) \ {d(v_Fq)
\over[G_+^0(\kktil)]^{-1} -2v_Fq } \
\psi^*_{+\alpha}(\kktil)\psi_{+\alpha}(\kktil)\cr  &  \simeq& \!\!\!\! -{\delta
E_0\over 2} (g_1^2(\ell)+g_2^2(\ell)
-g_1(\ell)g_2(\ell))\sum_{\alpha} \left\{\int_{-\dm(E_0(\ell)+E_0)}^{-E_0(\ell)} \!\!
+\int_{E_0(\ell)}^{\dm(E_0(\ell)+E_0)} \right\}
  \ {d(v_Fq) \over[G_+^0(\kktil)]^{-1} -2v_Fq } \
\psi^*_{+\alpha}(\kktil)\psi_{+\alpha}(\kktil)\cr
  & \simeq&  \!\!\!\!{1\over 4} (g_1^2(\ell)+g_2^2(\ell) -g_1(\ell)g_2(\ell))
 t(\ell)d\ell\sum_{\alpha} [G_+^0(\kktil)]^{-1}\
\psi^*_{+\alpha}(\kktil)\psi_{+\alpha}(\kktil)
\label{local}
\end{eqnarray}
to  leading order in $[G_+^0(\kktil)]^{-1}$. Here   $\ell_q =\ln  E_0/\mid v_F q\mid$,
$t(\ell)=1-2/(1+e^{\ell})$ and
$E_0(\ell_{N-3}) \to E_0(\ell)$ in the limit $d\ell \to 0$. The final
 expression  then becomes 
the outer shell contribution  of the renormalization coefficient $z^{-1}(\ell)$
of $[G_+^0]^{-1}$ for the inner shell state $k$ in the local approximation. The presence of the factor
$t(\ell)$ indicates that the above result is not  logarithmic over the entire range, especially at small 
$\ell$ where transients or  scaling deviations  exist.\cite{Bourbon91} These become vanishingly small at
large
$\ell$  and are   neglected to logarithmic accuracy.   A similar cascade of contractions applies to
the second term of (\ref{lambda}) with
$\lambda^-$  and this  leads to the same renormalization for the
one-particle propagator near $-k_F$.
\subsection{Umklapp contribution}
\label{sixu}
When umklapp scattering is present, the contraction $\delta S_u^{(j)}\equiv
\dm\langle
(S_{I,1})^2\big\vert_{g_3}\rangle_{\bar{0},c}$, that involves $g_3$ alone, generates a  three-particle
interaction at the step
$(j-2)d\ell$ with
$j\ge 3$: 

\begin{eqnarray}
  \delta S_u^{(j)} = \ (\pi v_F)^2{T^2\over L^2}\
\sum_{\{\kktil,\QQtil'\}}\sumslashD_{\QQtil'}\
\lambda_u^-(\ell_{j-3})\ G^0_-(\kktil'_1-\QQtil') &&\!\!\!\!\!\!\!
\psi^*_{+,\alpha_1}(\kktil_1+\QQtil)
\psi^*_{-,\alpha'_1}(\kktil_1-\QQtil-\QQtil'+\GGtil)
\psi^*_{-,\alpha'_2}(\kktil'_2+\QQtil'-\GGtil)\cr
 & \times & \psi_{-,\alpha_1}(\kktil_1)\psi_{
+,\alpha'_1}(\kktil'_1) \psi_{+,\alpha'_2}(\kktil'_2)\cr
+ \ (\pi v_F)^2{T^2\over L^2}\
\sum_{\{\kktil,\QQtil\}}\sumslashD_{\QQtil }\
\lambda_u^+(\ell_{j-3})\ G^0_+(\kktil_1+\QQtil)  && \!\!\!\!\!\!\!
\psi^*_{-,\alpha'_1}(\kktil-\QQtil')
\psi^*_{-,\alpha_1}(\kktil_1+\QQtil
+\QQtil'-\GGtil)\psi^*_{+,\alpha_2}(\kktil_2-\QQtil+\GGtil)\cr
 & \times &\psi_{+,\alpha'_1}(\kktil'_1) \psi_{
-,\alpha_1}(\kktil_1) \psi_{-,\alpha_2}(\kktil_2)
\label{sixthu}
\end{eqnarray}
where $\lambda_u^\pm(\ell_{j-3})=g_3^2(\ell_{j-3})$, $\QQtil=(2k_F +q,\omega_m)$
and $\GGtil=(4k_F,0)$. Here
$\sumslash_{\QQtil}$ ( $\sumslash_{\QQtil'}$) consists in summing all
boson frequencies and those
$q$ ($q'$)  values  of the  momentum transfer  for which $k_1+q$  
($k'_1-q'$) is in  the outershell
at step $(j-2)d\ell$. Following the example of the incommensurate case, the
next two contractions of the
first term of (\ref{sixthu}) will lead to corrections to the one-particle term at $+k_F$ with internal lines
in three successive outer shells and the external $k$ located at the top of the inner shell:  
\begin{eqnarray}
 \sum_{j=3}^N \langle \delta S'^{(j)}_{u,2} \rangle_{\bar{0},c}
\ =   \ (\pi v_F)^2{T^2\over L^2} \sum_{\alpha}\sum_{j=3}^N
 \ \sumslashD_{\kktil'}\!\!{}^{}\
\sumslashD_{\qtil}\,g_3^2(\ell_{j-3})
G^0_+(\kktil'+\QQtil)G_+^0(\kktil')G_-^0(\kktil+\QQtil-\GGtil)
\psi^*_{+\alpha}(\kktil)\psi_{+\alpha}(\kktil)
\label{nonlocalu}
\end{eqnarray}
 The explicit evaluation runs parallel to the incommensurate
case in (\ref{Green}-\ref{nonlocal}). Thus in the local approximation, the variation of
$g_3(\ell_{j-3})$  with
$j$ is neglected and the  value of the coupling is taken   at
$(N-1)d\ell\to \ell$. The outer shell umklapp
contribution to the single particle term at the inner shell state $k$ of the action then becomes 
\begin{eqnarray}
\sum_{j=3}^N \ \langle \delta S'^{(j)}_{u,2} \rangle_{\bar{0},c} \simeq {1\over
8} g_3^2(\ell) t(\ell)d\ell
\sum_{\alpha}[G_+^0(\kktil)]^{-1}
\psi^*_{+\alpha}(\kktil)\psi_{+\alpha}(\kktil).
\label{localu}
\end{eqnarray}
 This result, which adds to $z^{-1}(d\ell)$, corresponds to
the diagram of Figure~\ref{sixth}-e. Using the second term of (\ref{sixthu})
a similar expression holds for fermions at
$-k_F$.


\begin{thebibliography}{10}
\bibitem{note1}
The notation used here for the renormalization factors of $\mu_S$ differs from
  the one of Refs.~\cite{Bourbon91,Bourbon95}, from which the correspondance
  $z\leftrightarrow z^{-1}_1$, $z_1\leftrightarrow z_2$, $z_2\leftrightarrow
  z_3$, and $z_3\leftrightarrow z_4$ can be established.
\end{thebibliography}
\end{document}